\documentclass[a4paper,fleqn]{cas-sc}
\usepackage{amsmath}
\usepackage{graphicx}
\usepackage{epsf}
\usepackage{psfrag}
\usepackage{epsfig}
\usepackage{graphics}
\usepackage{amsfonts}
\usepackage{slashed}
\usepackage{footnote}
\usepackage[utf8]{inputenc}
\usepackage[T1]{fontenc}
\usepackage{dsfont}
\usepackage{subfigure}
\usepackage{subcaption}

\def\tsc#1{\csdef{#1}{\textsc{\lowercase{#1}}\xspace}}
\tsc{WGM}
\tsc{QE}

\begin{document}

\newcommand{\be}{\begin{equation}}
\newcommand{\ee}{\end{equation}}
\newcommand{\bq}{\begin{eqnarray}}
\newcommand{\eq}{\end{eqnarray}}

\newcommand{\Dslash}{\hbox{$\partial\!\!\!{\slash}$}}
\newcommand{\qslash}{\hbox{$q\!\!\!{\slash}$}}
\newcommand{\pslash}{\hbox{$p\!\!\!{\slash}$}}
\newcommand{\bslash}{\hbox{$b\!\!\!{\slash}$}}
\newcommand{\kslash}{\hbox{$k\!\!\!{\slash}$}}
\newcommand{\kbruto}{\hbox{$k \!\!\!{\slash}$}}
\newcommand{\pbruto}{\hbox{$p \!\!\!{\slash}$}}
\newcommand{\qbruto}{\hbox{$q \!\!\!{\slash}$}}
\newcommand{\lbruto}{\hbox{$l \!\!\!{\slash}$}}
\newcommand{\bbruto}{\hbox{$b \!\!\!{\slash}$}}
\newcommand{\parbruto}{\hbox{$\partial \!\!\!{\slash}$}}
\newcommand{\Abruto}{\hbox{$A \!\!\!{\slash}$}}
\newcommand{\bbbruto}{\hbox{$b_1 \!\!\!{\slash}$}}
\newcommand{\bbbbruto}{\hbox{$b_2 \!\!\!{\slash}$}}

\let\WriteBookmarks\relax
\def\floatpagepagefraction{1}
\def\textpagefraction{.001}

\shorttitle{}    

\shortauthors{}  

\title [mode = title]{Do anomalies break the momentum routing invariance?}  



%

\author[]{A. R. Vieira}






\affiliation[]{organization={Instituto de Ci\^encias Agr\'arias, Exatas e Biol\'ogicas de Iturama - ICAEBI, Universidade Federal do Tri\^angulo Mineiro - UFTM},
            city={Iturama},
            postcode={38280-000}, 
            state={Minas Gerais},
            country={Brasil}}




\begin{abstract}
The diagrammatic computation of anomalies is usually associated with the breaking of the momentum routing invariance. This is because 
the momentum routing is usually chosen to fulfill the desired Ward identity. In the case of the chiral anomaly, the momentum routing 
is chosen in order to fulfill the gauge Ward identity and break the chiral Ward identity. Although the chiral anomaly is physical 
because it is associated with the pion decay into two photons, this does not necessarily mean that the momentum routing invariance is 
broken because the momentum routing was chosen in the computation of the anomaly. The reciprocal is not true, {\it i. e.} anomalies 
do not imply in momentum routing invariance breaking. In this work, we show that if gauge invariance is assumed, the chiral and the
scale anomalies are independent of the momentum routing chosen and as a result they are momentum routing invariant. This idea is applied
to a QED with non-minimal CPT and Lorentz violation, where momentum routing invariance is used to find out what symmetry is broken in
the Ward identities.
\end{abstract}




\maketitle

\section{Introduction}

Symmetries are in the main core of physics. Our current knowledge of physical laws are built based on the concept of symmetry and symmetry breaking.
For instance, the form of particle interactions is determined by the symmetries of the model, Super-symmetric theories are built based on the Poincar\'e group 
and the very concept of an elementary particle, its mass and its spin are related to symmetry. Not to mention that the process of mass generation in the 
Standard Model of particles comes from the spontaneous symmetry breaking of a larger gauge symmetry in a smaller one. Over and above that, we desire that 
physical laws be unchanged under boosts and rotations, the existence of magnetic monopoles would reveal an expected symmetry of the Maxwell equations and the 
unbalanced asymmetry between matter and anti-matter in the universe is still an unanswered question. 

Symmetries are so important for field theories that historically they were taken for granted not only at the classical level but also at the quantum one. 
Therefore, the name anomaly was given to a quantum breaking of a classical symmetry as it was something unusual and not desired. At the same time, 
regularization schemes usually break symmetries of the theory and restoring counter-terms are then required in the process of renormalization. Thus, the 
question whether the anomaly is indeed physical or spurious, {\it i. e.} caused by the regularization scheme, is frequently raised. For instance, Lattice
regularization turns space-time discrete and this breaks Lorentz symmetry, among others like Super-symmetry and chiral symmetry \cite{Costa}, or cutoff 
regularization explicitly breaks gauge symmetry and it can be constructed to maintain this symmetry and the Lorentz one \cite{Cynolter0}. However, in both cases 
the breaking of the symmetries is an artifact of the regularization. On the other hand, anomalies are related to physical processes and therefore can be 
measured. Pioneer works revealed that the chiral anomaly is related to the neutral pion decay into two photons \cite{Jackiw}-\cite{Bardeen} and the trace 
anomaly of the Quantum Electrodynamics (QED) is related to the hadronic $R$ ratio \cite{Ellis}. Nowadays, there are numerous applications of the anomalies like 
in the quantum Hall effect of Weyl semi-metals \cite{Ahmad}, chiral magnetic effects for theories with chemical potentials \cite{Claudio}, form factors of 
particle processes in effective field theories \cite{Hao}, glueball mass spectrum \cite{glueball} or even the relation of chiral anomalous processes with a 
quark anomalous magnetic moment \cite{{Hao},{Xing}}. The application of the chiral anomaly in solid state physics is even more diverse than in particle
physics. Since one of its first applications to Weyl fermions in a crystal \cite{Nielsen}, the chiral anomaly was shown to affect the magneto-transport of Weyl
semimetals \cite{PRB,Science,Nature1,Nature2} and observations of negative magneto-resistance for different materials support the existence of this anomaly \cite
{Science}-\cite{Nature2}. For a review on the Weyl and Dirac semimetals and the applications of the chiral anomaly in these materials see \cite{Review}.

Since its discovery, the chiral anomaly was believed to be affected by momentum routing ambiguities among other anomalies like 
supersymmetric ones \cite{Prereg, McKeon}. Some authors consider this approach as a pre-regularization \cite{Prereg} since the 
momentum routing is appropriately chosen for the fulfillment of a Ward identity before the regularization of any integral. However, 
since the anomaly has been measured, it seems that there was a preferred momentum routing choice in the computation of the anomaly in the previous theoretical predictions prior this measurement. The theoretical puzzle in this case is solved thanks to experiments. 
Nevertheless, it remains the question: does the anomaly necessarily imply in the momentum routing invariance breaking? In the case of 
supersymmetric anomalies or anomalies in frameworks like the Lorentz and CPT- violating Standard Model, there would be no experimental 
data available to solve a theoretical issue related to the ambiguity in the momentum routing choice in the anomaly computation. One can instead alternatively resort to momentum routing invariance since there is also no observation that shows that momentum routing invariance is broken in Feynman diagrams. In this work, we present examples that show that the diagrammatic computation of anomalies, for a specific momentum routing, does not necessarily imply that momentum routing invariance of the Feynman diagrams is broken because it is also possible to derive the anomalies for a general momentum routing.  In 
particular, we consider the examples of the chiral anomaly (we adopt this nomenclature although it is also known as the axial anomaly, 
the Adler-Bardeen-Bell-Jackiw anomaly or even the triangle anomaly) for chiral abelian gauge theories and the scale anomaly for QED. Since it is explicitly shown that these anomalies do not depend on the momentum routing, it is possible to find out the correct result of them without appealing to experiments to 
solve the issue of ambiguity in the momentum routing choice.

The paper is divided as follows: in section \ref{sIR}, we present a summary of implicit regularization. In section 
\ref{sMRI}, we discuss some examples of momentum routing invariance of the Feynman diagrams and its relation to gauge symmetry. In 
section \ref{CAZ}, we present a simple approach to compute the chiral anomaly. In section \ref{ABJ}, we present a non-trivial approach 
to compute the chiral anomaly for general momentum routings of the internal lines. In section \ref{sSA}, we compute the scale anomaly 
of QED for a general momentum routing and show that, as in the chiral anomaly example, the result of the anomaly is independent of the 
momentum routing. These ideas are applied to a novel situation in section \ref{sTLV}, where the chiral anomaly is studied in a QED
with non-minimal CPT and Lorentz violation. Finally, we present the conclusions in section \ref{sC}.

\section{Outline of Implicit Regularization}
\label{sIR}

We apply the implicit regularization scheme \cite{IR} to treat the integrals which appear in the amplitudes of the next sections. Some 
regularization schemes like dimensional regularization \cite{{DR},{Bollini}}, and its extensions like dimensional reduction 
\cite{{Siegel}, {Siegel2}}, although widely used is a regularization scheme which is already gauge and momentum routing invariant and 
some of the following discussions would be trivial. On the other hand, implicit regularization allows the computation of the 
amplitudes for a general momentum routing and it is suitable specially for amplitudes with dimensional specific objects such as 
$\gamma^5$ matrices because the number of dimensions do not have to be changed. The results of the regularized integrals presented in 
the appendix can be simply obtained in conventional dimensional regularization if the number of dimensions is changed to $d$ and the 
surface terms are made zero.

Implicit regularization is generally applied to theories with dimension specific objects, like $\gamma^5$ matrices and Levi-Civita 
symbols. Also, since it is
a scheme that does not have any restriction (except for theories with non-linear vertices such as the Sine-Gordon model) and it does not break symmetries of the 
theory, it is usually used for computation of anomalies. A recent computation for a general momentum routing concerns gravitational anomalies in two dimensions 
\cite{Orimar}. Other scenarios with Lorentz violation, like in the Bumblebee model \cite{Ricardo}, or chiral models \cite{Ricardo2} deal directly 
with $\gamma^5$ matrices and comparison with other regularization techniques is performed \cite{{Ricardo},{Ricardo2},{Adriano3}}. Let us make a brief review of 
the method in four dimensions. In this scheme, we assume that the integrals are regularized by an implicit regulator $\Lambda$ in order to allow algebraic  
operations within the integrands. We then use the following identity 
\begin{equation}
\int_k\frac{1}{(k+p)^2-m^2} =\int_k\frac{1}{k^2-m^2}
 -\int_k\frac{(p^2+2p\cdot k)}{(k^2-m^2)[(k+p)^2-m^2]},
\label{2.1}
\end{equation}
where $\int_k\equiv\int^\Lambda\frac{d^4 k}{(2\pi)^4}$, to separate basic divergent integrals (BDI's) from the finite part. These BDI's are defined as follows
\begin{equation}
I^{\mu_1 \cdots \mu_{2n}}_{log}(m^2)\equiv \int_k \frac{k^{\mu_1}\cdots k^{\mu_{2n}}}{(k^2-m^2)^{2+n}}
\end{equation}
and
\begin{equation}
I^{\mu_1 \cdots \mu_{2n}}_{quad}(m^2)\equiv \int_k \frac{k^{\mu_1}\cdots k^{\mu_{2n}}}{(k^2-m^2)^{1+n}}.
\end{equation}

The BDI's with Lorentz indexes can be judiciously combined as differences between integrals with the same superficial degree 
of divergence, according to the equations below, which define surface terms  \footnote{The Lorentz indexes between brackets stand for 
permutations, {\it i.e.}$A^{\{\alpha_1\cdots\alpha_n\}}B^{\{\beta_1\cdots\beta_n\}}=A^{\alpha_1\cdots\alpha_{n}}
B^{\beta_1\cdots\beta_n}$ + sum over permutations between the two sets of indexes $\alpha_1\cdots\alpha_{n}$ and 
$\beta_1\cdots\beta_n$. For instance, $g^{\{\mu\nu}g^{\alpha\beta\}}=g^{\mu\nu}g^{\alpha\beta}
+g^{\mu\alpha}g^{\nu\beta}+g^{\mu\beta}g^{\nu\alpha}$.}:
\begin{eqnarray}
\Upsilon^{\mu \nu}_{2w}=  g^{\mu \nu}I_{2w}(m^2)-2(2-w)I^{\mu \nu}_{2w}(m^2) 
\equiv \upsilon_{2w}g^{\mu \nu},
\label{dif1}\\
\nonumber\\
\Xi^{\mu \nu \alpha \beta}_{2w}=  g^{\{ \mu \nu} g^{ \alpha \beta \}}I_{2w}(m^2)
 -4(3-w)(2-w)I^{\mu \nu \alpha \beta }_{2w}(m^2)\equiv\nonumber\\
\equiv  \xi_{2w}(g^{\mu \nu} g^{\alpha \beta}+g^{\mu \alpha} g^{\nu \beta}+g^{\mu \beta} g^{\nu \alpha}).
\label{dif2}
\end{eqnarray} 

In the expressions above, $2w$ is the degree of divergence of the integrals and we adopt the notation such that indexes $0$ and $2$ 
mean $log$ and $quad$, respectively. Surface terms can be conveniently written as integrals of total derivatives, as presented below

\begin{eqnarray}
\upsilon_{2w}g^{\mu \nu}= \int_k\frac{\partial}{\partial k_{\nu}}\frac{k^{\mu}}{(k^2-m^2)^{2-w}}, \nonumber \\
\label{ts1}
\end{eqnarray}
\begin{eqnarray}
(\xi_{2w}-v_{2w})(g^{\mu \nu} g^{\alpha \beta}+g^{\mu \alpha} g^{\nu \beta}+g^{\mu \beta} g^{\nu \alpha})= \int_k\frac{\partial}{\partial 
k_{\nu}}\frac{2(2-w)k^{\mu} k^{\alpha} k^{
\beta}}{(k^2-m^2)^{3-w}}.
\label{ts2}
\end{eqnarray}

We see that equations (\ref{dif1})-(\ref{dif2}) are undetermined because they are differences between divergent quantities. Each 
regularization scheme gives a different value for these terms. However, as physics should not depend on the scheme applied and neither
on the arbitrary surface term, we leave these terms to be arbitrary until the end of the calculation and then we fix them by symmetry 
constraints or phenomenology \cite{JackiwFU}.

Of course the same idea can be applied for any dimension of space-time and for higher loops. Equation (\ref{2.1}) is used recursively 
until the divergent piece is separated from the finite one. This procedure makes the finite integrals hard to compute due to the 
number of $k$'s in the numerator. A simpler alternative to this approach is presented in \cite{Bruno}, where the Feynman 
parametrization is applied before separating the BDI's. However, in order to apply this 
simpler approach we have to assume that the integrals are momentum routing invariant in order to apply the Feynman parametrization in 
the divergent integral. This would compromise our proof in the examples of the next sections once that we want to show that anomalies 
are momentum routing invariant. Also, eq. (\ref{2.1}) is not the only possible equation to be used since the implicit regulator was 
assumed to allow the use of other identities as well.

\section{Momentum Routing Invariance of Feynman diagrams}
\label{sMRI}

The choice of momentum routing of the internal lines is not usually an issue in the computation of loop diagrams. Since dimensional 
regularization was successfully applied to gauge field theories to prove their renormalizability, it is common to apply this scheme to 
perform loop computations and it allows for shifts in the integral momentum even for divergent integrals. So, dimensional 
regularization is already momentum routing invariant besides being gauge invariant as it is well known. The issue arises for example 
in the chiral anomaly computation since dimensional regularization can not be employed due to the presence of the $\gamma_5$ matrix, 
which is not well defined for $d$ dimensions. This issue can be addressed by redefining it to $d$ dimensions or by splitting the loop 
momentum in a parallel and a perpendicular pieces \cite{Peskin}. Therefore, the choice of the internal momentum routing is compatible 
with the correct result of the chiral anomaly. This might lead to the conclusion that since anomalies are physical, the momentum 
routing invariance is broken in anomalous diagrams because the momentum routing had to be chosen in order to fulfill the gauge Ward 
identity and break the chiral Ward identity by the correct value of the chiral anomaly. However, as we are going to present in the 
examples below, there is no preferred momentum routing even for anomalous diagrams.

\begin{figure}[!h]
\centering
\includegraphics[clip, trim=0mm 30mm 0mm 15mm , width=0.5\textwidth]{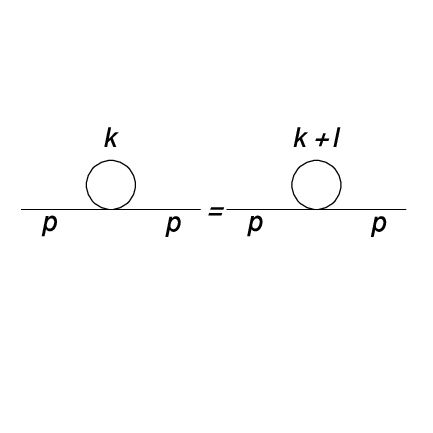}
\caption{Momentum routing invariance of a 1-loop Feynman diagram of the scalar theory $\phi^4$.}
\label{fig1}
\end{figure}

Let us consider the simple example of a 1-loop Feynman diagram of a $\phi^4$ scalar theory presented in figure \ref{fig1}. We see that 
there is a freedom of adding an arbitrary momentum routing $l$ in the internal line because this is also compatible with momentum 
conservation at the vertex. The Feynman rules lead to the following equality of integrals if we require momentum routing invariance:
\begin{equation}
\int\frac{d^4k}{(2\pi )^4}\frac{1}{(k^2-m^2)}=\int\frac{d^4k}{(2\pi )^4}\frac{1}{[(k+l)^2-m^2]}. 
\label{eqMRIS}
\end{equation} 

If this 1-loop diagram is part of a scattering process, physics should not depend on the way we label the internal lines. The integral 
in the left hand side of eq.(\ref{eqMRIS}) is quadratic divergent and making a shift in the integral momentum by adding a $l$ 
changes its result \cite{Pugh}. The shifts do not change the result of the integrals only when they are finite or logarithmic 
divergent. Eq. (\ref{eqMRIS}) is actually a requirement of momentum routing invariance for this amplitude. Both of these integrals 
require a regularization scheme. For instance, dimensional regularization \cite{{DR},{Bollini}} is a scheme that allows for shifts in 
the integrand even if the integral is divergent. So, the identity in eq. (\ref{eqMRIS}) is trivially fulfilled. We can, alternatively, 
assume that an implicit regulator exists and compute the integrals, as presented in section \ref{sIR}. The regularized integrals are 
found in the appendix. When doing this computation a surface term appear and if we require momentum routing invariance we find out 
that 
\begin{align}
I_{quad}(m^2)&=I_{quad}(m^2)-l^2\upsilon_0\nonumber\\
&\Rightarrow l^2\upsilon_0=0.
\label{eqSD}
\end{align}

Since the momentum routing $l$ can be any momentum, the solution is to make the arbitrary surface term $\upsilon_0$ equal to zero to 
assure momentum routing invariance and make the scattering process independent of this. The renormalization of scalar theories was 
performed beyond 1-loop order and it was shown that observables such as the beta function would depend on the arbitrary surface terms 
if we do not require momentum routing invariance of the Feynman diagrams \cite{Adriano0}. Besides the beta function, the amplitudes 
are measurable in a decay or in a scattering process and, therefore, they should not depend on the way we label the momentum in 
the internal lines of a loop. Momentum routing invariance is also important in the sense that physical amplitudes should not depend on 
this labeling of momenta.

\begin{figure}[ht]
\centering
\subfigure[]{
\centering
\includegraphics[trim=0mm 85mm 70mm 90mm, scale=0.4]{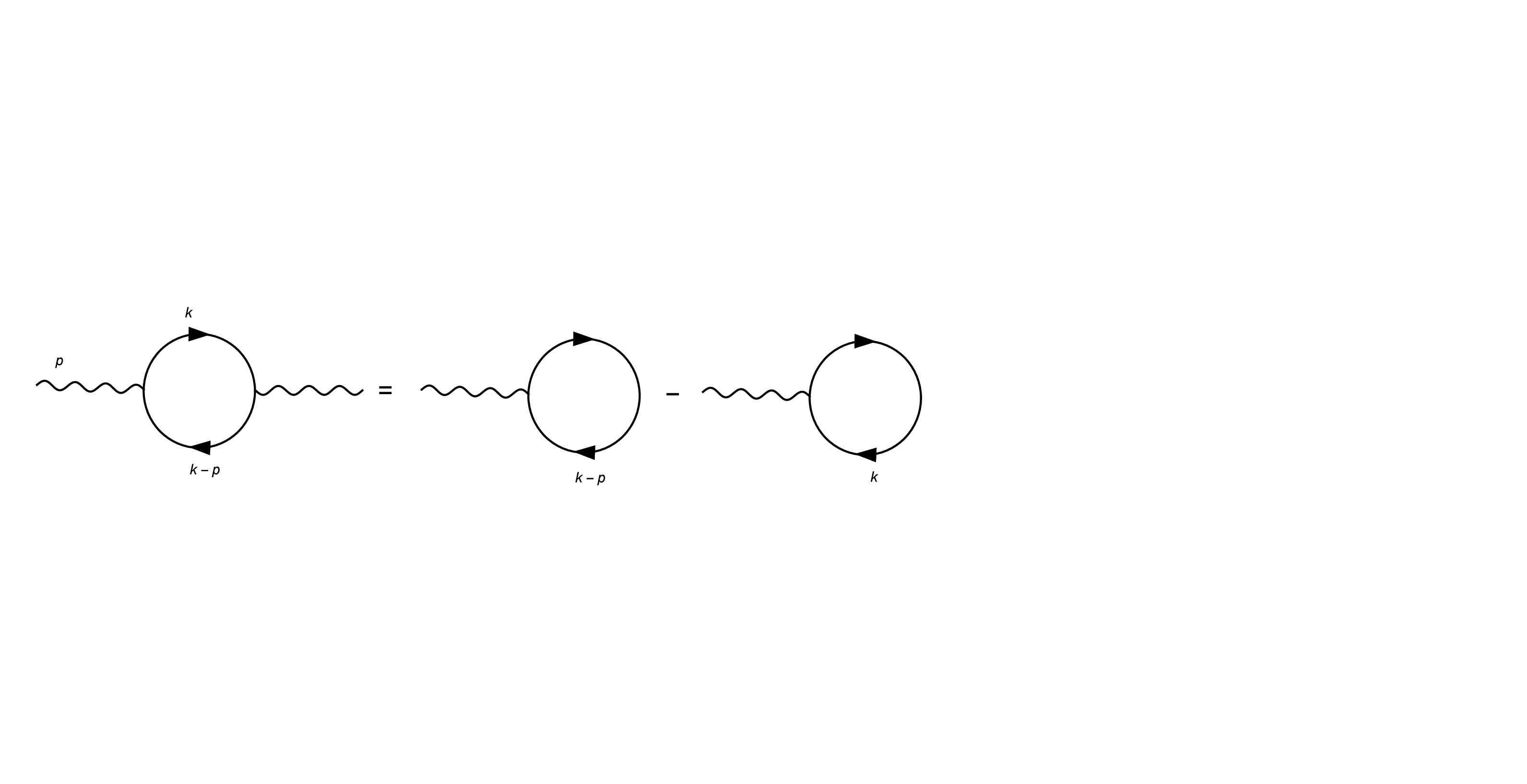}}
\subfigure[]{\includegraphics[trim=0mm 40mm 0mm 80mm , width=0.8\textwidth]{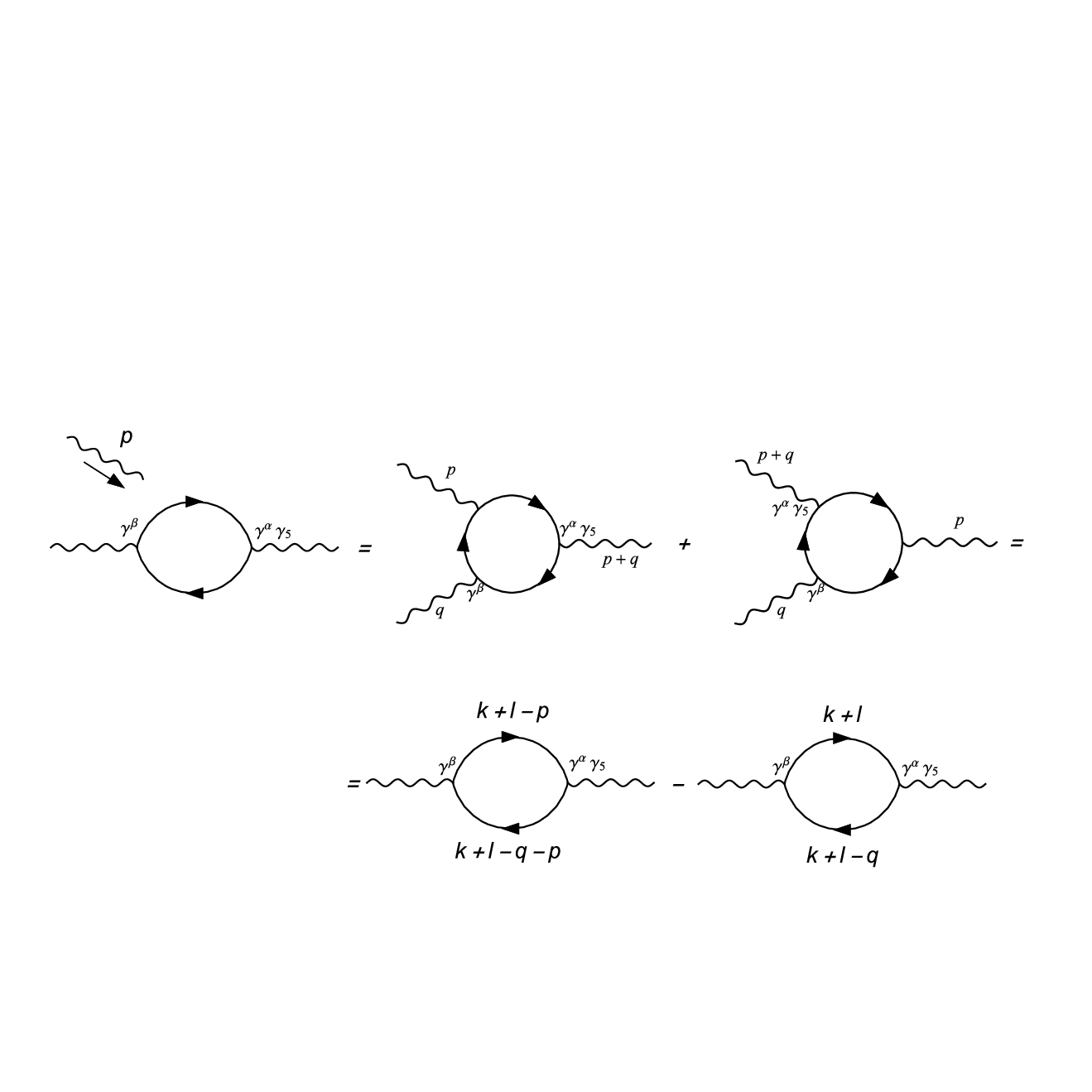}}
\caption{(a) The gauge and momentum routing invariance relation for the vacuum polarization tensor. (b) The diagrammatic gauge and 
momentum routing invariance relation for a Ward identity of a three leg diagram in a chiral gauge theory. The photon momenta $q$ and 
$p$ are ingoing and the photon momentum $p+q$ is outgoing.}
\label{fig2}
\end{figure}

In abelian gauge field theories and in chiral ones, there is an additional reason that leads to momentum routing invariance. Gauge 
symmetry is diagrammatically related to momentum routing invariance in a regularization independent way. As a first example, we
can consider the gauge Ward identity of the vacuum polarization tensor. Applying the simple algebra 
$\slashed{p}=\slashed{k}-m-(\slashed{k}-\slashed{p}-m)$, this identity can be trivially rewritten as a difference between
two tadpoles with different momentum routings as it is shown below and presented in figure \ref{fig2}:
\begin{equation}
p_{\mu}\Pi^{\mu\nu}(p)=-e^2\int_k Tr\left[\slashed{p}\frac{1}{\slashed{k}-\slashed{p}-m}\gamma^{\nu}\frac{1}{\slashed{k}-m}\right]=
-e^2\int_k Tr\left[\gamma^{\nu}\frac{1}{\slashed{k}-\slashed{p}-m}\right]+e^2\int_k Tr\left[\gamma^{\nu}\frac{1}{\slashed{k}-m}
\right],
\label{eqMRIVP}
\end{equation}
where $\int_k\equiv \int \frac{d^4 k}{(2\pi)^4}$.

We see in eq. (\ref{eqMRIVP}) that if there is gauge symmetry, there is automatically momentum routing invariance and vice versa.
In this way, the momentum routing invariance besides being related to the momentum conservation in the
vertices of the Feynman diagrams, is directly related to gauge symmetry in a regularization independent way. However, some 
regularizations can confirm this relation. For instance, since shifts are allowed in dimensional regularization, the right hand side 
of eq. (\ref{eqMRIVP}) is trivially zero. Also, if the trace is taken and the divergent integrals are evaluated (see the appendix)  
with the use of implicit regularization, the result of eq. (\ref{eqMRIVP}) is:
\begin{equation}
p_{\mu}\Pi^{\mu\nu}(p)=4e^2(p^{\nu}\upsilon_2+p^2 p^{\nu}(\xi_0-2\upsilon_0)),
\label{eqWIE}
\end{equation}
where the surface terms $\upsilon_2$, $\upsilon_0$ and $\xi_0$ are defined in section \ref{sIR}.

Thus, if gauge invariance is required in eq. (\ref{eqWIE}), we find conditions on the surface terms, {\it i. e.} $\upsilon_2=0$ and 
$\xi_0=2\upsilon_0$. As we are going to see shortly, this are the same conditions we would obtain if the vacuum polarization tensor
was explicitly computed and then contracted with the momentum $p_{\mu}$.

As another example, we can consider a two point diagram where we insert an external momentum photon leg. The gauge Ward identity is 
obtained by inserting the photon leg wherever is possible in the diagram \cite{Peskin}. In the case of a two point diagram, there are 
only two possibilities. After performing the simple algebra
$\slashed{p}=\slashed{k}+\slashed{l}-m-(\slashed{k}+\slashed{l}-\slashed{p}-m)$, we can split both diagrams and find out that the 
gauge Ward identity is simply the difference between two diagrams with different momentum routing as we can see in figure \ref{fig2} 
and algebraically as below:
\begin{align}
p_{\lambda}A^{\lambda\mu\alpha}=&\int_k Tr \left[\frac{1}{\slashed{k}+\slashed{l}-m}\slashed{p}\frac{1}{\slashed{k}+\slashed{l}-\slashed{p}-m}\gamma^{\beta}
\frac{1}{\slashed{k}+\slashed{l}-\slashed{p}-\slashed{q}-m}\gamma^{\alpha}\gamma^{5}\right]\nonumber\\
&+\int_k Tr \left[\frac{1}{\slashed{k}+\slashed{l}-m}\gamma^{\beta}\frac{1}{\slashed{k}+\slashed{l}-\slashed{q}-m}\slashed{p}\frac{1}{\slashed{k}+\slashed
{l}-\slashed{p}-\slashed{q}-m}\gamma^{\alpha}\gamma^{5}\right] \nonumber\\
=&\int_k Tr \left[\frac{1}{\slashed{k}+\slashed{l}-\slashed{p}-m}\gamma^{\beta}\frac{1}{\slashed{k}+\slashed{l}-\slashed{q}-\slashed{p}-m}\gamma^{\alpha}
\gamma^{5}\right]\nonumber\\
&-\int_k Tr \left[\frac{1}{\slashed{k}+\slashed{l}-m}\gamma^{\beta}\frac{1}{\slashed{k}+\slashed{l}-\slashed{q}-m}\gamma^{\alpha}\gamma^{5}
\right],
\label{bolha2}
\end{align}
where $l$ is an arbitrary momentum routing.

So, if we require gauge symmetry, we have automatically momentum routing invariance. This proof can be generalized for an arbitrary 
number of legs and loops \cite{MRICGT}. The gauge and momentum routing invariance relation was also observed in other models like Wess 
-Zumino \cite{Adriano} and the Lorentz-violating QED \cite{MRILV}. Nevertheless, the gauge and momentum routing invariance relation
is broken by a mass term if the theory contains fermions of different species and massive gauge bosons \cite{Bruno}. This would be the 
case if the gauge boson changes the flavor in the vertex as the $W^{\pm}$ bosons.

It is also possible to find the requirements for momentum routing invariance in QED. Let us consider the vacuum polarization tensor
whose computation in implicit regularization is given by
\bq
\centering
i\Pi^{\mu\nu}(p)&=\frac{4}{3}e^2(p^2g^{\mu\nu}-p^{\mu}p^{\nu})I_{log}(m^2)-4e^2\upsilon_2g^{\mu\nu}+\frac{4}{3}e^2(p^2g^{\mu\nu}-p^{\mu}p^{\nu})\upsilon_0
-\frac{4}{3}e^2(p^2g^{\mu\nu}+2p^{\mu}p^{\nu})(\xi_0-2\upsilon_0)+\nonumber\\
&-\frac{i}{2\pi^2}e^2(p^2g^{\mu\nu}-p^{\mu}p^{\nu})(Z_1-Z_2),
\label{eqpi}
\eq
where the finite integrals are defined according to the equations below
\begin{align}
&Z_n=Z_n(p^2,m^2)\equiv\int_0^1 dx x^n\ln \left(\frac{D(x)}{m^2}\right),\label{finiteint1}\\
&\iota_n=\int_0^1 dx \frac{x^n(1-x)}{D(x)}, \label{finiteint1}\\
&\textnormal{and}\nonumber\\
&D(x)=m^2-p^2 x(1-x).
\label{finiteint3}
\end{align}

Notice that if we require gauge invariance using the Ward identity $p_{\mu}\Pi^{\mu\nu}(p)=0$, we find that the quadratic surface term 
$\upsilon_2$ must be zero and that the logarithmic surface terms must obey the relation $\xi_0=2\upsilon_0$. Now, if we add an 
arbitrary routing $l$, we find out that eq. (\ref{eqpi}) acquires additional surface terms and both finite and divergent pieces are 
unaffected:
\begin{align}
\centering
i\Pi^{\mu\nu}(l,p)=&\frac{4}{3}e^2(p^2g^{\mu\nu}-p^{\mu}p^{\nu})I_{log}(m^2)-4e^2\upsilon_2 g^{\mu\nu}-\nonumber\\
&-4e^2\xi_0\left( l^2 g^{\mu  \nu }+g^{\mu  \nu } l\cdot p+\frac{p^2}{3} g^{\mu  \nu }+ l^{\nu } p^{\mu }+ l^{\mu } p^{\nu }+2 l^{\mu } l^{\nu }+\frac{2 p^{\mu } p^{\nu }}{3}\right)+ \nonumber\\
&+e^2\upsilon_0 \left(4 g^{\mu  \nu } \left(2 l^2+2 l\cdot p+p^2\right)+4 \left(2 l^{\mu }+p^{\mu }\right) \left(2 l^{\nu }+p^{\nu }\right)\right)
-\frac{i}{2\pi^2}e^2(p^2g^{\mu\nu}-p^{\mu}p^{\nu})(Z_1-Z_2),
\label{eqpi2}
\end{align}

\begin{figure}[!h]
\centering
\includegraphics[trim=0mm 60mm 0mm 60mm , width=0.7\textwidth]{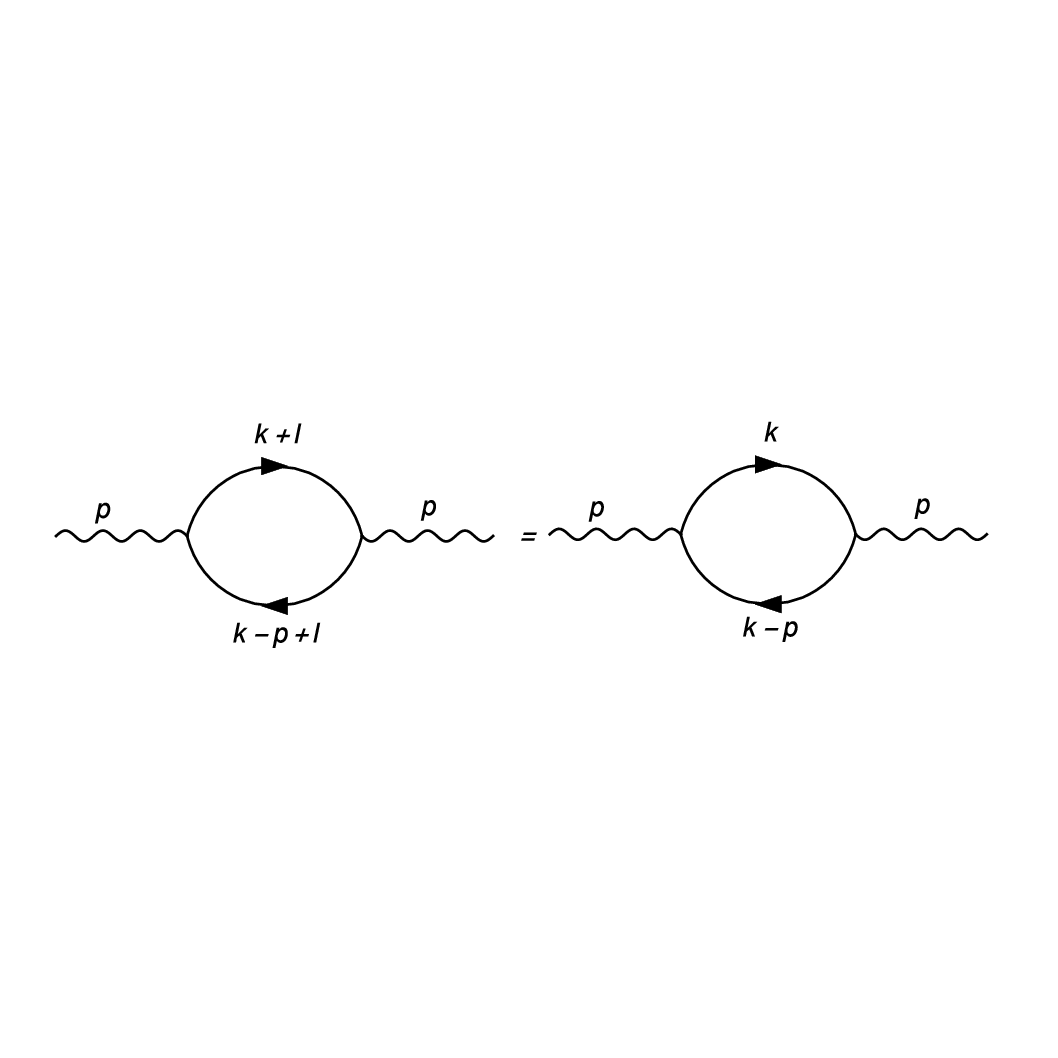}
\caption{Momentum routing invariance of the vacuum polarization tensor.}
\label{fig3}
\end{figure}

Notice that we recover eq. (\ref{eqpi}) from eq. (\ref{eqpi2}) for $l=0$. The momentum routing invariance condition is the same 
required before for the scalar loop and it is presented in figure \ref{fig3}. Since the divergent and the finite pieces of the 
amplitude are physical, the only terms that remains depends on the surface terms and the general momentum routing $l$ as expected. 
Again, if this 1-loop diagram is part of a scattering process, physics should not depend on the way we assign momenta routings in the 
internal lines. Thus, we have the condition
\begin{align}
\centering
&\Pi^{\mu\nu}(l,p)=\Pi^{\mu\nu}(0,p) \nonumber\\
&-4 (\xi_0-2\upsilon_0 ) \left(l^2 g^{\mu  \nu }+g^{\mu  \nu } l\cdot p+l^{\nu } p^{\mu }+l^{\mu } p^{\nu }+2 l^{\mu } l^{\nu }\right)=0
\label{eqMRIC}
\end{align}

Since the general momentum routing $l$ can be any value, the only possible solution to eq. (\ref{eqMRIC}) is to make 
$\xi_0=2\upsilon_0$. This is the same condition that assures gauge invariance, as it was expected because the gauge-momentum routing 
invariance relation is independent of regularization.

The Feynman diagrams should be unchanged under the transformation $k\rightarrow k+l$, where $k$ is the integrated momentum of an 
internal line and $l$ is a generic momentum routing. It would be tempting to call momentum routing invariance a symmetry of the theory 
because of that. However, this is not the case because this momentum transformation is not associated with a transformation in the 
fields of the theory that makes the action invariant. We can instead call it a "symmetry" or a feature of the Feynman diagrams. 

\section{Chiral anomaly: the usual and simpler computation }
\label{CAZ}

\begin{figure}[!h]
\centering
\includegraphics[trim=0mm 30mm 0mm 30mm, scale=0.65]{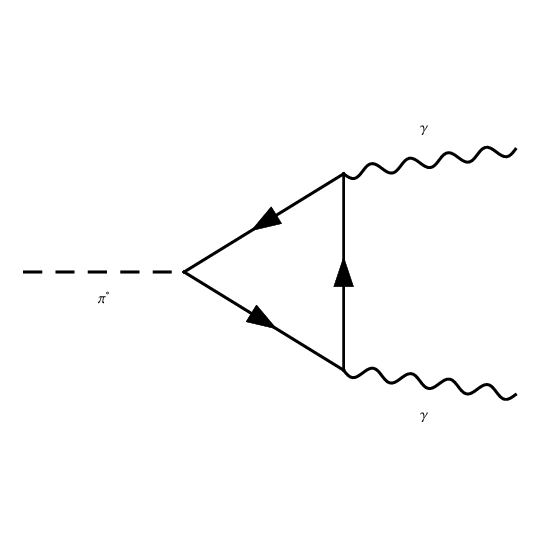}
\caption{The neutral pion decay into two photons.}
\label{figp}
\end{figure}

The chiral current  $J^{\mu}_5=\bar{\psi} \gamma^{\mu}\gamma^5\psi$ is conserved for 
fermionic massless theories and a loop correction that comes from a $VVA$ triangle diagram breaks this symmetry at the quantum level.
The lowest order process occur at 1-loop with a linear divergent amplitude and, therefore, a regularization scheme is required.
Regardless of the theoretical prediction, from an experimental point of view, this diagram is related to the amplitude of the pion 
decay into two photons, among other physical processes. The pion decay process is illustrated in figure \ref{figp}. In this sense, the 
chiral anomaly is physical because it can be measured in a particle process. The relation between the amplitude 
$\mathcal{M}_{\pi^0\rightarrow \gamma+\gamma}$ of the pion decay and the triangle diagram is presented below:\\
\begin{align}
&\mathcal{M}_{\pi^0\rightarrow \gamma+\gamma}= \epsilon_{\nu}(p)\epsilon'_{\mu}(q)T^{\mu\nu}(p,q)\\
&T^{\mu\nu}(p,q)=\frac{\mu^2-(p+q)^2}{F \mu^2} (p_{\alpha}+q_{\alpha})T^{\mu\nu\alpha}(p,q)\\
&T^{\mu\nu\alpha}(p,q)=-ie^2 \int d^4 x\ d^4 y e^{-i p\ x}e^{-i q\ y}\langle0\mid T\ J^{\alpha}_5(0) J^{\mu}(x)J^{\nu}(y) \mid 0\rangle,
\end{align}\\
where $\epsilon_{\nu}(p)$ and $\epsilon'_{\mu}(q)$ are different photon polarization vectors, $J^{\mu}=\bar{\psi} \gamma^{\mu}\psi$ is 
the vector current, $\mu$ is the pion mass and $F$ is a form factor.

\begin{figure}[!h]
\centering
\includegraphics[trim=0mm 180mm 0mm 40mm , width=0.8\textwidth]{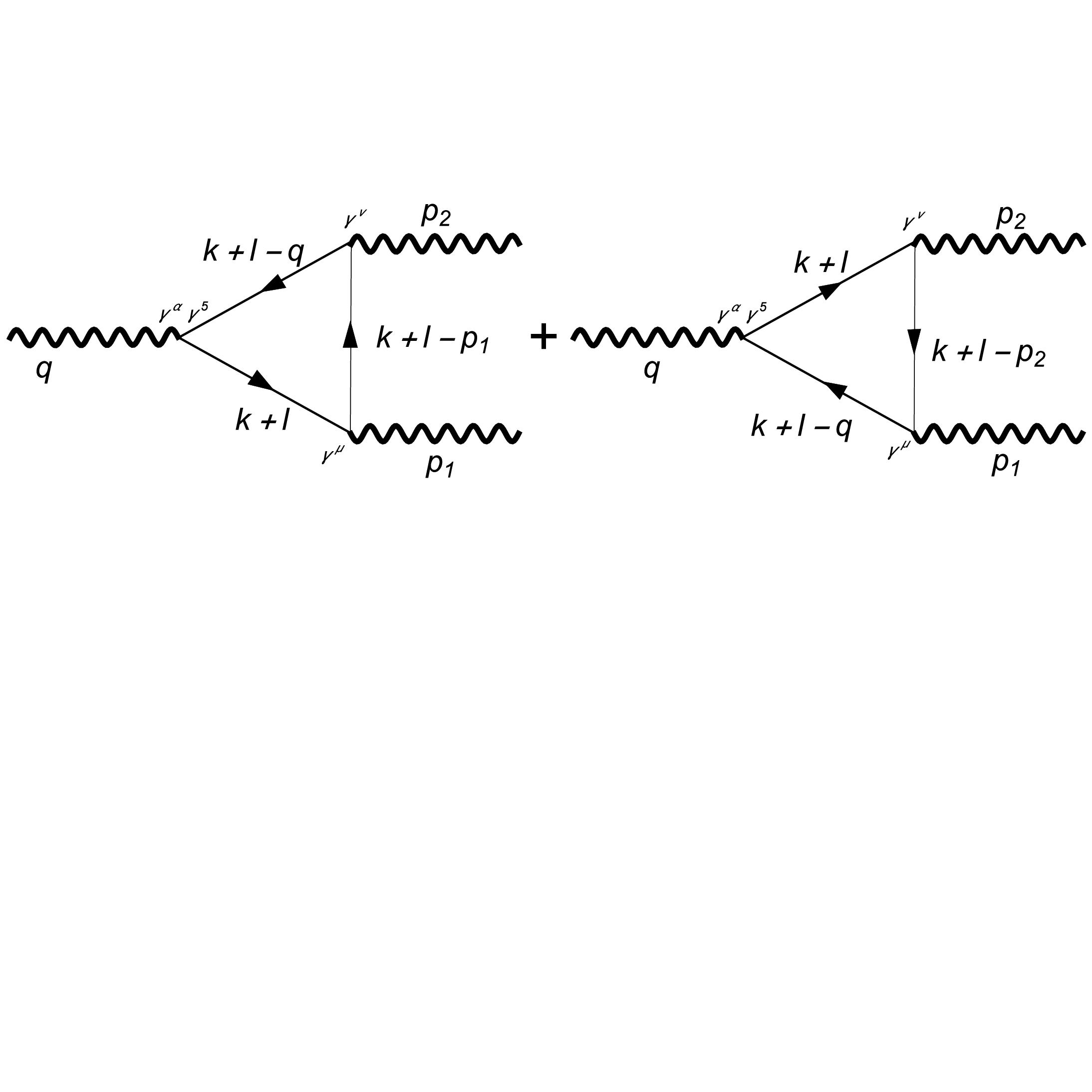}
\caption{Triangle diagrams which contribute to the chiral anomaly. The internal lines have an arbitrary momentum route $l$.}
\label{fig0}
\end{figure}

The computation of the chiral anomaly found in textbooks or in the original papers usually chooses the momentum routing in order to fulfill the gauge Ward 
identity and get the correct result for the chiral anomaly. In order to illustrate this, we follow \cite{{Zee2}, {Bertlmann}}. The amplitude of the two 
triangle diagrams shifted by a general momentum $l$ and depicted in figure \ref{fig0} is given by:
\begin{equation}
T^{\lambda\mu\nu}(l, p_1, p_2)=-\int \frac{d^4 k}{(2\pi)^4}Tr\left[\gamma^{\alpha}\gamma^{5}\frac{1}{\slashed{k}-\slashed{q}+\slashed{l}}\gamma^{\nu}
 \frac{1}{\slashed{k}-\slashed{p}_1+\slashed{l}}\gamma^{\mu}\frac{1}{\slashed{k}+\slashed{l}}\right]+(\mu\leftrightarrow\nu ,p_1\leftrightarrow p_2).
\label{eqRDC}
\end{equation} 

In order to compute eq. (\ref{eqRDC}) one relies on the fact that the surface term generated by the shift $k\rightarrow k+l$ can be computed with the use of 
the Taylor series and the Green's theorem, which lead us to 
\begin{equation}
\int d^4k [f(k+l)-f(k)]=\lim_{P\rightarrow \infty} i l_{\mu}\left(\frac{P^{\mu}}{P}\right)f(P)(2\pi^2 P^3).
\label{eqSTC}
\end{equation}

However, this is not the only possible result for a surface term. As presented in section \ref{sIR}, surface terms can be any number 
because they are differences between two divergent integrals. They are also regularization dependent. So, the scheme applied to 
compute the integral, symmetric integration in case of eq. (\ref{eqSTC}), assigns a particular value to the surface term.

To proceed, we define
\begin{equation}
f(k)=Tr\left[\gamma^{\alpha}\gamma^{5}\frac{1}{\slashed{k}-\slashed{q}}\gamma^{\nu}
 \frac{1}{\slashed{k}-\slashed{p}_1}\gamma^{\mu}\frac{1}{\slashed{k}}\right],
\end{equation}
which lead us to the result \[ \lim_{P\rightarrow \infty}f(P)=\frac{Tr[\gamma^{\alpha}\gamma^5\slashed{P}\gamma^{\nu}\slashed{P}\gamma^{\mu}\slashed{P}]}{P^6}=
\frac{4iP_{\sigma}\epsilon^{\sigma\nu\mu\alpha}}{P^4}\] and 
\begin{equation}
T^{\alpha\mu\nu}(l, p_1, p_2)-T^{\alpha\mu\nu}(p_1, p_2)=\frac{1}{2\pi^2}\lim_{P\rightarrow \infty}l^{\omega}\frac{P_{\omega}P_{\sigma}}{P^2}\epsilon^{
\sigma\nu\mu\alpha}+(\mu\leftrightarrow\nu, p_1\leftrightarrow p_2)=\frac{1}{8\pi^2} l_{\sigma}\epsilon^{\sigma\nu\mu\alpha}+(\mu\leftrightarrow\nu, p_1
\leftrightarrow p_2),
\label{eqCA1}
\end{equation}
where we assumed an average over the surface of the sphere to use that $\frac{P_{\mu}P_{\nu}}{P^2}=\frac{1}{4}g^{\mu\nu}$.

We then write the general momentum routing $l$ as $l=\alpha(p_1+p_2)+\beta(p_1-p_2)$ and insert it in eq. (\ref{eqCA1}) to find out that
\begin{equation}
T^{\alpha\mu\nu}(l, p_1, p_2)=T^{\alpha\mu\nu}(p_1, p_2)+\frac{\beta}{4\pi^2} (p_{1\sigma}-p_{2\sigma})\epsilon^{\sigma\nu\mu\alpha},
\label{eqWI1}
\end{equation}
where we apply the Ward identity $p_{1\mu}T^{\alpha\mu\nu}(l, p_1, p_2)=0$ to require gauge invariance.

Alternatively, one can compute from eq. (\ref{eqRDC}) for $l=0$ the identity
\begin{equation}
p_{1\mu}T^{\alpha\mu\nu}(p_1, p_2)=-\frac{1}{8\pi^2} p_{1\mu}p_{2\sigma}\epsilon^{\alpha\nu\mu\sigma},
\label{eqWI2}
\end{equation}
from where we find out with the use of eq. (\ref{eqWI1}) that $\beta=\frac{1}{2}$.

Finally, we can compute the chiral anomaly by contracting $q_{\alpha}$ with eq. (\ref{eqWI1}):
\begin{align}
&q_{\alpha}T^{\alpha\mu\nu}(l, p_1, p_2)=q_{\alpha}T^{\alpha\mu\nu}(p_1, p_2)+\frac{1}{8\pi^2}q_{\alpha} (p_{1\sigma}-p_{2\sigma})\epsilon^{\sigma\nu\mu\alpha} \nonumber\\
&q_{\alpha}T^{\alpha\mu\nu}(l, p_1, p_2)=-\frac{1}{2\pi^2}p_{1\alpha} p_{2\sigma}\epsilon^{\sigma\nu\mu\alpha},
\label{eqWI3}
\end{align}
where we used the result $q_{\alpha}T^{\alpha\mu\nu}(p_1, p_2)=-\frac{1}{4\pi^2}p_{1\alpha} p_{2\sigma}\epsilon^{\sigma\nu\mu\alpha}$ that can be computed
from eq. (\ref{eqRDC}) for $l=0$.

Since the routing was chosen to fulfill the gauge Ward identities and get the correct result of the chiral anomaly, it looks that the momentum routing 
invariance is now violated because we had to choose a particular momentum in order to do this computation. In other words, if $\beta=\frac{1}{2}$ we
have $l=(\alpha+1/2)p_1+(\alpha-1/2)p_2$ and it remains only one free parameter. As we are going to show in the next section, instead of choosing the momentum 
routing one can alternatively choose the arbitrary surface term.


\section{Chiral anomaly: a momentum routing invariant computation}
\label{ABJ}

The literature on the chiral anomaly presents its computation by several different approaches \cite{Elias}-\cite{Wu}. As it is well 
known, dimensional regularization is not suitable for loop diagrams with $\gamma^5$ matrices because these matrices are not well 
defined in $d$ dimensions. This issue can be addressed if the loop momenta is split in a parallel and a perpendicular piece \cite
{Peskin}. There is also several other approaches addressed to deal with this $\gamma^5$ matrix issue in dimensional regularization 
besides the dimensional reduction \cite{{Siegel}, {Siegel2}} regularization scheme, including new proposals like the rightmost 
position approach \cite{{Tsai}, {Tsai2}} and the variant of the original Kreimer prescription with a constructively-defined $\gamma^5$ 
\cite{Chen}. An overview on the various regularization schemes applied in the diagrammatic anomaly computation can be found in \cite
{Bertlmann}. Besides the diagrammatic computations, there are also the Fujikawa approach based on the path integral measure 
transformation \cite{Fujikawa, Fujikawa2} and the approach that applies differential geometry \cite{Zee}. These last approaches are 
not diagrammatic computations and therefore they do not depend on the momentum routing of the internal lines.

In this section we derive the chiral anomaly applying the implicit regularization presented in section \ref{sIR}. We show that the 
regularization dependent content is contained in the surface term. Following the idea presented in \cite{JackiwFU}, the arbitrariness 
inherent to some perturbative calculations in quantum field theory should be fixed based on the symmetries of the model that we want 
to preserve. For instance, in the neutral pion decay into two photons, we must preserve the vector Ward identities and, consequently, 
the axial one is violated \cite{{Jackiw},{Adler}}. On the other hand, in the Standard Model the chiral coupling with gauge fields 
refers to fermion-number conservation and the axial identity must be enforced \cite{tHooft}.

The approach presented in this section leads to a momentum routing independent result for the chiral anomaly, {\it i. e.} the 
computation of this anomaly is perform with arbitrary routings of the internal momenta. As the result for the scalar theory in eq. (
\ref{eqSD}) and the one for QED in eq. (\ref{eqMRIC}), the momentum routing multiply the arbitrary surface terms. Therefore, is 
possible to choose a value for the surface term instead of choosing the momentum routing as in section \ref{CAZ}. 

Another issue found in triangle diagrams refers to objects like $Tr[\gamma ^{\mu }\gamma ^{\beta }\gamma ^{\nu }\gamma ^{\xi }\gamma 
^{\alpha }\gamma ^{\lambda }\gamma ^5]$ that appears in the amplitudes. The following identity can be used to reduce the number of 
Dirac $\gamma$ matrices
\be
\gamma ^{\mu }\gamma ^{\beta }\gamma ^{\nu }= g^{\mu \beta} \gamma^{\nu}+g^{\nu \beta} \gamma^{\mu}-g^{\mu \nu} \gamma^{\beta}-i 
\epsilon^{\mu \beta \nu \rho}\gamma_{\rho}\gamma^5.
\label{gamma}
\ee

Using eq. (\ref{gamma}),  the identities $Tr[\gamma ^{\mu }\gamma ^{\beta }\gamma ^{\nu }\gamma ^{\xi }\gamma ^5]=4i\epsilon^{\mu \beta \nu \xi}$ 
and $\gamma^5\gamma^{\rho}\gamma^5=-\gamma^{\rho}$ we find the result below which is the one that most textbooks work with
\cite{Wu}
\be
Tr[\gamma ^{\mu }\gamma ^{\beta }\gamma ^{\nu }\gamma ^{\xi }\gamma ^{\alpha}\gamma ^{\lambda }\gamma ^5]=4i(g^{\beta \mu}\epsilon^{\nu 
\xi \alpha \lambda}+g^{\beta \nu}\epsilon^{\mu \xi \alpha \lambda}-g^{\mu \nu}\epsilon^{\beta \xi \alpha \lambda}
-g^{\lambda \alpha}\epsilon^{\mu \beta \nu \xi}+g^{\xi \lambda}\epsilon^{\mu \beta \nu \alpha}-g^{\xi \alpha}\epsilon^{\mu \beta \nu 
\lambda}).
\label{trace}
\ee

However, it is completely arbitrary which three $\gamma$ matrices we pick up to apply equation (\ref{gamma}). A different choice would 
give the result of equation (\ref{trace}) with Lorentz indexes permuted. Besides, the equation $\{\gamma^5,\gamma_{\mu}\}=0$ should be avoided inside 
a divergent integral \cite{{MRICGT}, {Adriano2}} since this operation seems to fix a value for the surface term. This point
of view is also shared by other works \cite{{Cynolter}, {Luciana}, {Perez}}. Therefore, we adopt a version of eq. (\ref{trace}) which contains all 
possible Lorentz structures available. This version reads
\begin{align}
\frac{-i}{4}Tr[\gamma ^{\mu }\gamma ^{\nu }\gamma ^{\alpha }\gamma ^{\beta }\gamma ^{\gamma }\gamma ^{\delta }\gamma ^5]&=-g^{\alpha 
\beta } \epsilon ^{\gamma \delta \mu \nu }+g^{\alpha \gamma } \epsilon ^{\beta \delta \mu \nu }-g^{\alpha \delta } \epsilon ^{\beta 
\gamma \mu \nu }-g^{\alpha \mu } \epsilon ^{\beta \gamma \delta \nu }+g^{\alpha \nu } \epsilon ^{\beta \gamma \delta \mu }-g^{\beta 
\gamma } \epsilon ^{\alpha \delta \mu \nu }+\nonumber\\&+g^{\beta \delta } \epsilon ^{\alpha \gamma \mu \nu }+g^{\beta \mu } \epsilon ^{
\alpha \gamma \delta \nu }-g^{\beta \nu } \epsilon ^{\alpha \gamma \delta \mu }-g^{\gamma \delta } \epsilon ^{\alpha \beta \mu \nu }-g^{
\gamma \mu } \epsilon ^{\alpha \beta \delta \nu }+g^{\gamma \nu } \epsilon ^{\alpha \beta \delta \mu }+\nonumber\\&+g^{\delta \mu } 
\epsilon ^{\alpha \beta \gamma \nu }-g^{\delta \nu } \epsilon ^{\alpha \beta \gamma \mu }-g^{\mu \nu } \epsilon ^{\alpha \beta \gamma 
\delta },
\label{traco2}
\end{align}
and it can also be obtained if we replace $\gamma^5$ by its definition, 
$\gamma^5=\frac{i}{4!}\epsilon^{\mu\nu\alpha\beta}\gamma_{\mu}\gamma_{\nu}\gamma_{\alpha}\gamma_{\beta}$, and take the trace.

Equation (\ref{traco2}) has already been used in similar computations of previous works \cite{{Cynolter}, {Perez}, {Wu}, {Luciana}}. It was also generalized to 
be included in $(n+1)-$point Green functions \cite{Luciana}.

\begin{figure}[!h]
\centering
\includegraphics[trim=0mm 180mm 0mm 40mm , width=0.8\textwidth]{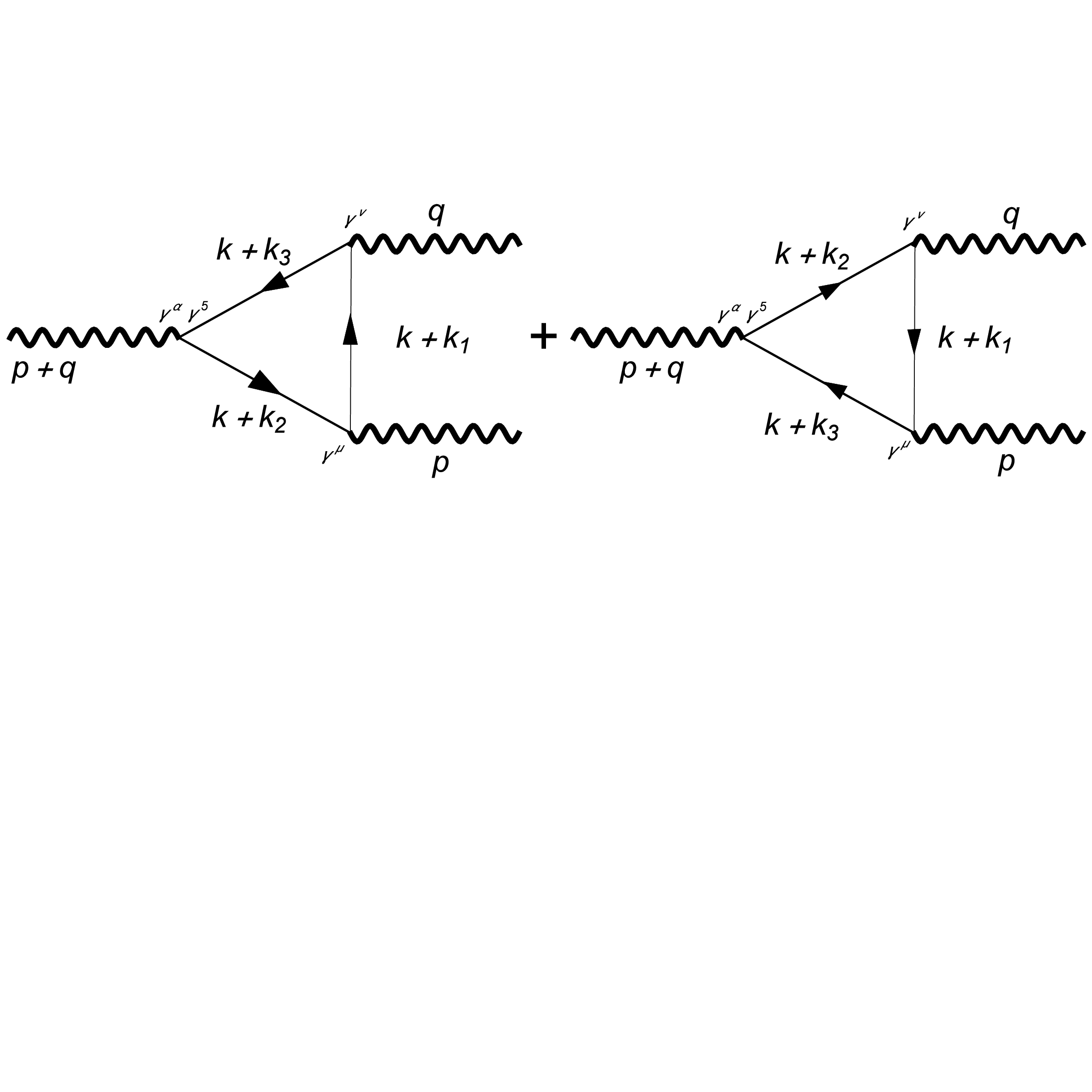}
\caption{Triangle diagrams which contribute to the chiral anomaly. We label the internal lines with arbitrary momentum routing.}
\label{triangle}
\end{figure}
The amplitude of the Feynman diagrams of figure \ref{triangle} is given by
\be
T_{\mu\nu\alpha}=-i\int_k Tr\left[\gamma_{\mu}\frac{i}{\slashed{k}+\slashed{k}_1-m}\gamma_{\nu}\frac{i}{\slashed{k}+\slashed{k}_2-m}
\gamma_{\alpha}\gamma^5 \frac{i}{\slashed{k}+\slashed{k}_3-m}\right]+(\mu \leftrightarrow \nu, p\leftrightarrow q).
\label{AVV}
\ee
where the arbitrary routing $k_i$ obeys the following relations due to energy-momentum conservation at each vertex
\begin{align}
k_2-k_3=& p+q,\nonumber\\
k_1-k_3=& p,\nonumber\\
k_2-k_1=& q.
\label{routing}
\end{align}

Equations (\ref{routing}) allow us to parametrize the routing  $k_i$ as
\begin{align}
k_1=& \alpha p+(\beta-1) q,\nonumber\\
k_2=& \alpha p+\beta q,\nonumber\\
k_3=& (\alpha-1) p+(\beta-1) q,
\label{routing2}
\end{align}
where $\alpha$ and $\beta$ are arbitrary real numbers which map the freedom we have in choosing the momentum routing of internal lines, since any
combination of $q$ and $p$ is possible as long as it is according to vertex momentum conservation in eqs. (\ref{routing}).
Equations (\ref{routing}) and (\ref{routing2}) for the other diagram are obtained by changing $ p\leftrightarrow q$.

After taking the trace using equation (\ref{traco2}), we apply the implicit regularization scheme in order to regularize the integrals coming from equation (\ref{AVV}). The result is 
\be
T_{\mu\nu\alpha}= 4i\upsilon_0 (\alpha-\beta-1)\epsilon_{\mu \nu \alpha \beta}(q-p)^{\beta}+T^{finite}_{\mu\nu\alpha},
\label{resultado}
\ee
where $\upsilon_0$ is a surface term defined in section \ref{sIR} and $T^{finite}_{\mu\nu\alpha}$ is the finite part of the amplitude 
whose evaluation we perform in the appendix.

We then apply the respective external momentum in equation (\ref{resultado}) in order to obtain the Ward identities:
\begin{align}
&p_{\mu}T^{\mu \nu \alpha}=-4i\upsilon_0 (\alpha-\beta-1)\epsilon^{\alpha \nu \beta \lambda}p_{\beta}q_{\lambda},\nonumber\\
&q_{\nu}T^{\mu \nu \alpha}=4i\upsilon_0 (\alpha-\beta-1)\epsilon^{\alpha \mu \beta \lambda}p_{\beta}q_{\lambda},\nonumber\\
&(p+q)_{\alpha}T^{\mu \nu \alpha}=2m T_5^{\mu \nu}+8i\upsilon_0 (\alpha-\beta-1)\epsilon^{\mu \nu \beta \lambda}p_{\beta}q_{\lambda}-\frac
{1}{2\pi^2}\epsilon^{\mu \nu \beta \lambda}p_{\beta}q_{\lambda},
\label{WI}
\end{align}
where $T_5^{\mu \nu}$ is the usual vector-vector-pseudo-scalar triangle.

The number $\upsilon_0 (\alpha-\beta-1)$ is arbitrary since $\upsilon_0$ is a difference of
two infinities and $\alpha$ and $\beta$ are any real numbers that we have freedom in choosing as long as equations (\ref{routing}) 
representing the energy-momentum conservation hold. We can parametrize this arbitrariness in a single parameter $a$ redefining
\begin{equation}
4i\upsilon_0(\alpha-\beta-1)\equiv\frac{1}{4\pi^2}(1+a).
\label{eqad}
\end{equation}

With the use of (\ref{eqad}), the Ward identities in equations (\ref{WI}) reads

\begin{align}
&p_{\mu}T^{\mu \nu \alpha}=-\frac{1}{4\pi^2}(1+a)\epsilon^{\alpha \nu \beta \lambda}p_{\beta}q_{\lambda},\nonumber\\
&q_{\nu}T^{\mu \nu \alpha}=\frac{1}{4\pi^2}(1+a)\epsilon^{\alpha \mu \beta \lambda}p_{\beta}q_{\lambda},\nonumber\\
&(p+q)_{\alpha}T^{\mu \nu \alpha}=2m T_5^{\mu \nu}+\frac{1}{2\pi^2}a\epsilon^{\mu \nu \beta \lambda}p_{\beta}q_{\lambda}.
\label{WI22}
\end{align}
\noindent
From now, we will focus only in the massless theory since we would like to discuss just the quantum symmetry breaking term. 

If we want to maintain gauge invariance, we choose $a=-1$ and automatically the axial identity is violated by a quantity equal to
$-\frac{1}{2\pi^2}\epsilon^{\mu \nu \beta \lambda}p_{\beta}q_{\lambda}$, as the result obtained in the previous section. On the other hand, if we want to 
maintain chiral symmetry at the quantum level, we choose $a=0$ and the vectorial identities are violated. The choice $a=-1$ sets the surface term $\upsilon_0$ 
to zero. Since the surface term is zero to assure gauge invariance and it multiplies the arbitrary momentum routing, any value of the real numbers $\alpha$ and 
$\beta$ are possible, which makes the result momentum routing invariant. On the other hand, if gauge invariance is broken ($a=0$), there is a relation between 
the surface term and the momentum routing parameters $\alpha$ and $\beta$, which does not fix the momentum routing to a specific value but as in section \ref
{CAZ} the momentum routing now is not as general as possible because it only depends on the parameter $\alpha$. Furthermore, we notice that $a=-1$ is also 
compatible with $(\alpha-\beta-1)= 0$ and $\upsilon_0\neq 0$. In this case, the momentum routing is related according to the eq. $\alpha=\beta+1$ and it is
also possible to have gauge invariance for this specific momentum routings.

As presented in the previous section, the usual and simpler approach chooses a specific internal momentum in order to violate chiral symmetry and
preserve the gauge one. Nevertheless, the implicit regularization approach and the trace symmetrization allowed us to find out that the result of the chiral
anomaly is valid for any momentum routing if the arbitrary surface term $\upsilon_0$ is zero to assure gauge invariance. Notice, however, that is also possible
to have gauge invariance for the specific momentum routing $\alpha=\beta+1$ and $\upsilon_0\neq 0$. Therefore, the chiral anomaly may imply in the momentum routing invariance violation but not necessarily. It is also possible to have the correct result for the chiral anomaly compatible with momentum routing
invariance.

It is noteworthy that choosing a value for the surface term instead of choosing the momentum routing is more appealing. The surface 
terms can be any number because they are differences between two infinities. If computed with different regularizations, they lead to
different results. On the other hand, the general momentum routing is associated with the momentum conservation in the vertices and 
it is directly related to gauge invariance in a diagrammatic way, {\it i. e.} it does not depend on the regularization, although some regularizations like the dimensional and the implicit ones can confirm this relation.

\section{Scale anomaly computation for a general momentum routing $l$}
\label{sSA}

The scale anomaly was computed in the past for scalar theories \cite{Callan}, for QED \cite{{Ellis}, {LoopReg}} and also for non-abelian gauge theories \cite
{Collins}. Unlike the chiral anomaly, there are less debates on the scale anomaly concerning regularization schemes and different approaches. The first 
computations for scalar fields \cite{Callan} demonstrated with perturbation theory that scale anomalies arise due to the renormalization process. The further 
investigation for QED used the gauge invariant vacuum polarization tensor already known at that time and showed that the trace anomaly was directly related to 
particle processes \cite{Ellis}. That original computation was confirmed with the use of loop regularization \cite{LoopReg}. The terms that break scale 
symmetry at the quantum level are proportional to the beta function. So, besides being related to observables such as the hadronic $R$ ratio \cite{Ellis}, it 
is also related to the renormalization group functions. We would expect therefore that the scale anomalous term does not depend on the momentum routing. Before 
we proceed to the computation of the diagrams, we derive below the classical breaking terms and the Ward identity of the dilatation current.

Field theories are said to be scale invariant if they are unchanged under the following scale transformations:

\be
x'= e^{\epsilon} x
\label{eq0}
\ee

and

\be
\Phi'(x')= e^{-\epsilon d_{\Phi}} \Phi(x),
\label{eq1}
\ee
where $\epsilon$ is a scale parameter, $\Phi$ is a generic field and $d_{\Phi}$ is its scale dimension. Thus, these particular type of transformations
belong to the conformal group.


Kinetic terms in Lagrangians contain only fields or derivatives, both transform according to scale transformations, so that these 
kinetic terms have scale dimension equal to $4$. For instance, the fermion kinetic term in QED is 
$\bar{\psi}\gamma^{\mu}\partial_{\mu}\psi$ and its scale dimension is $4$ since the scale dimension of the fermion field and the 
derivative is $3/2$ and $1$, respectively, according to equation (\ref{eq1}). Interaction terms, in turn, have scale 
dimension equal to $4$ in the case of renormalizable quantum field theories because the couplings in these interaction terms are 
dimensionless. This would not be the case of non-renormalizable field theories because the couplings in these theories have ordinary 
dimension but have no scale dimension. The same thing happens for mass terms in lagrangians. Masses have no scale dimension and do not 
transform according to scale transformations.  A fermion mass term has scale dimension equal to $3$, for instance. If the term in 
the Lagrangian has scale dimension equal to $4$, the action is scale invariant because, when applying eq. (\ref{eq1}), the change 
caused by the transformation is $e^{-4\epsilon}$ and this is compensated when the change $e^{4\epsilon}$ in the volume element $d^{4}x$
is acquired according to equation (\ref{eq0}).

The current associated with the scale invariant transformations is the dilatation current and it is equal to the trace of the energy 
momentum-tensor. In QED, it is possible to build a symmetric energy-momentum tensor, known as the Belinfante tensor, whose trace leads 
to the classical violation of the dilatation current:
\be
\partial_{\mu}\mathcal{J}^{\mu}=\mathcal{T}^{\mu}_{\ \ \mu}=m\bar{\psi}\psi,
\ee
where $\mathcal{T}^{\mu}_{\ \ \mu}$ is the trace of the symmetric energy-momentum tensor.

Besides the mass terms, the quantum corrections also break scale symmetry because a renormalization scale is introduced in the 
renormalization process. For instance, in dimensional regularization, the scale parameter $\mu^{\epsilon-D}$ multiplies the divergent 
integrals whose dimension we changed for $D$. If one applies the scale transformations of eqs. (\ref{eq0}) and (\ref{eq1}), it seems 
that the Ward identity of the dilatation current is given by
\be
-i G_{\mathcal{T}}(0,p_1,p_2,...,p_{n-1})=\left(n(d_{\Phi}-4)+4-\sum_k^{n-1}p_k\frac{\partial}{\partial p_k}\right)G(p_1,p_2,...,p_{n-1}),
\label{eqIW}
\ee
where $G_{\mathcal{T}}(0,p_1,p_2,...,p_{n-1})$ and $G(p_1,p_2,...,p_{n-1})$ are $(n-1)$-point Green functions in the momentum space of 
the respective coordinate space $n-$point Green functions, $G_{\mathcal{T}}(y,x_1,x_2,...,x_{n})=\langle 
\mathcal{T}^{\mu}_{\mu}(y)\Phi(x_1)\Phi(x_2)...\Phi(x_{n})\rangle$ and $G(x_1,x_2,...,x_{n})=\langle 
\Phi(x_1)\Phi(x_2)...\Phi(x_{n})\rangle$. Note that the lower index $\mathcal{T}$ is to emphasize that one of the fields in the 
expectation value is the energy-momentum tensor. However, the derivation of eq. (\ref{eqIW}) assumes $\mathcal{D}\phi=\mathcal{D}\phi'$
in a scale transformation of the functional integral and this is not true for quantum corrections \cite{Fujikawa, Fujikawa2}. As we 
are going to see, the identity (\ref{eqIW}) is incomplete and acquires an anomalous term.

For $n=2$ and $\Phi$ being the vector field, eq. (\ref{eqIW}) reads
\bq
&-i G^{\mu\nu}_{\mathcal{T}}(0,p)=\left(n(d_{\Phi}-4)+4-p^{\lambda}\frac{\partial}{\partial p^{\lambda}}\right)G^{\mu\nu}(p), \nonumber\\
&-i G^{\mu\nu}_{\mathcal{T}}(0,p)=\left(-2-p^{\lambda}\frac{\partial}{\partial p^{\lambda}}\right)G^{\mu\nu}(p).
\label{eqIWG}
\eq

It is also possible to rewrite (\ref{eqIWG}) in terms of the one-particle-irreducible (1PI) function $\Gamma^{\sigma\lambda}(p)$, related to the legged 2-point function via the equation $G_{\mu\nu}(p)=D_{\mu\sigma}(p)\Gamma^{\sigma\lambda}(p)D_{\lambda\nu}(p)$, being $D_{\mu\nu}(p)=-\frac{ig_{\mu\nu}}{p^2}$ 
the photon propagator. The Ward identity in eq. (\ref{eqIWG}) can be written as

\be
-i \Gamma^{\mu\nu}_{\mathcal{T}}(0,p)=\left(2-p^{\lambda}\frac{\partial}{\partial p^{\lambda}}\right)\Gamma^{\mu\nu}(p).
\ee

\begin{figure}[!h]
\center
\subfigure[]{\includegraphics[trim=0mm 75mm 0mm 75mm,scale=0.55]{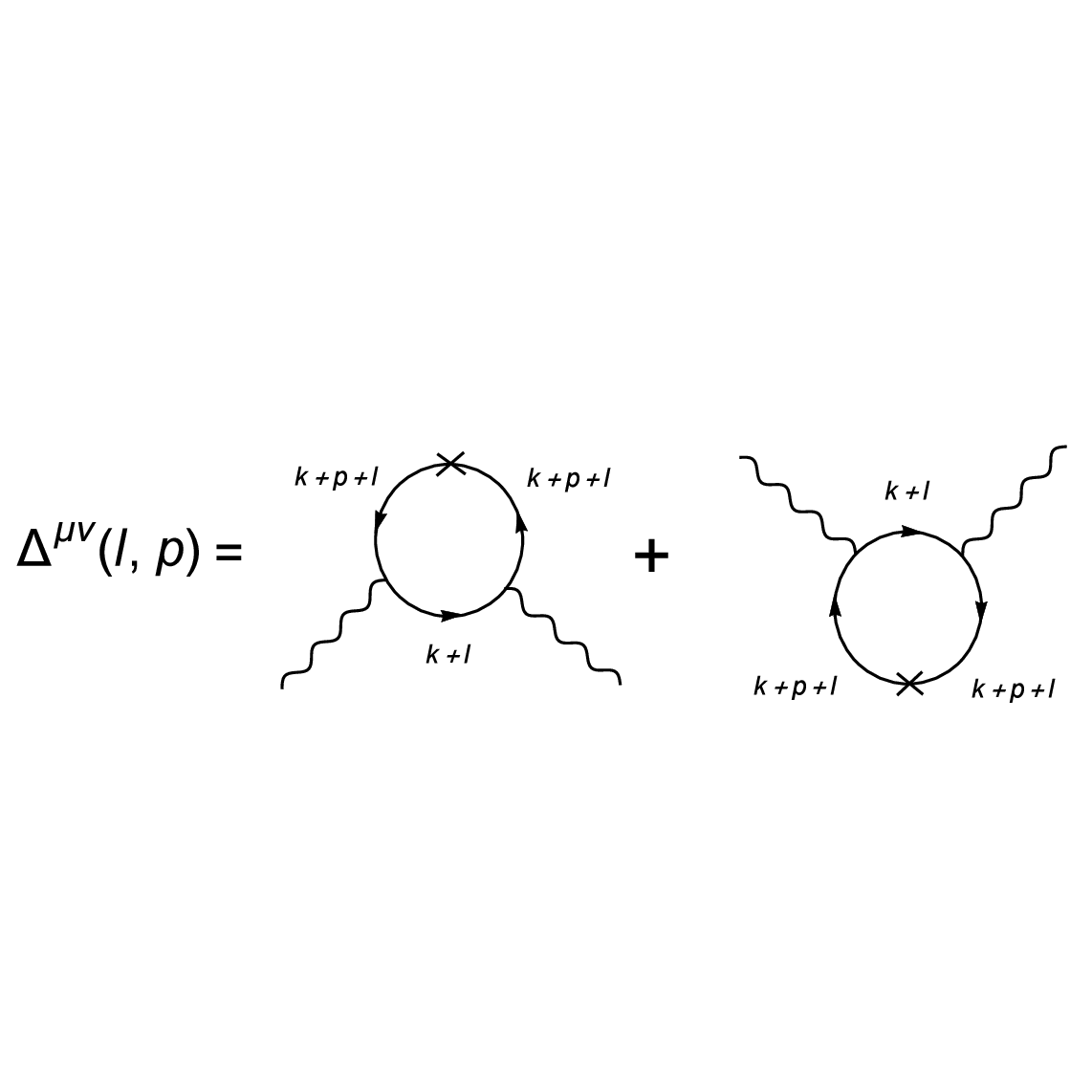}}
\qquad
\subfigure[]{\includegraphics[trim=0mm 55mm 0mm 55mm,scale=0.55]{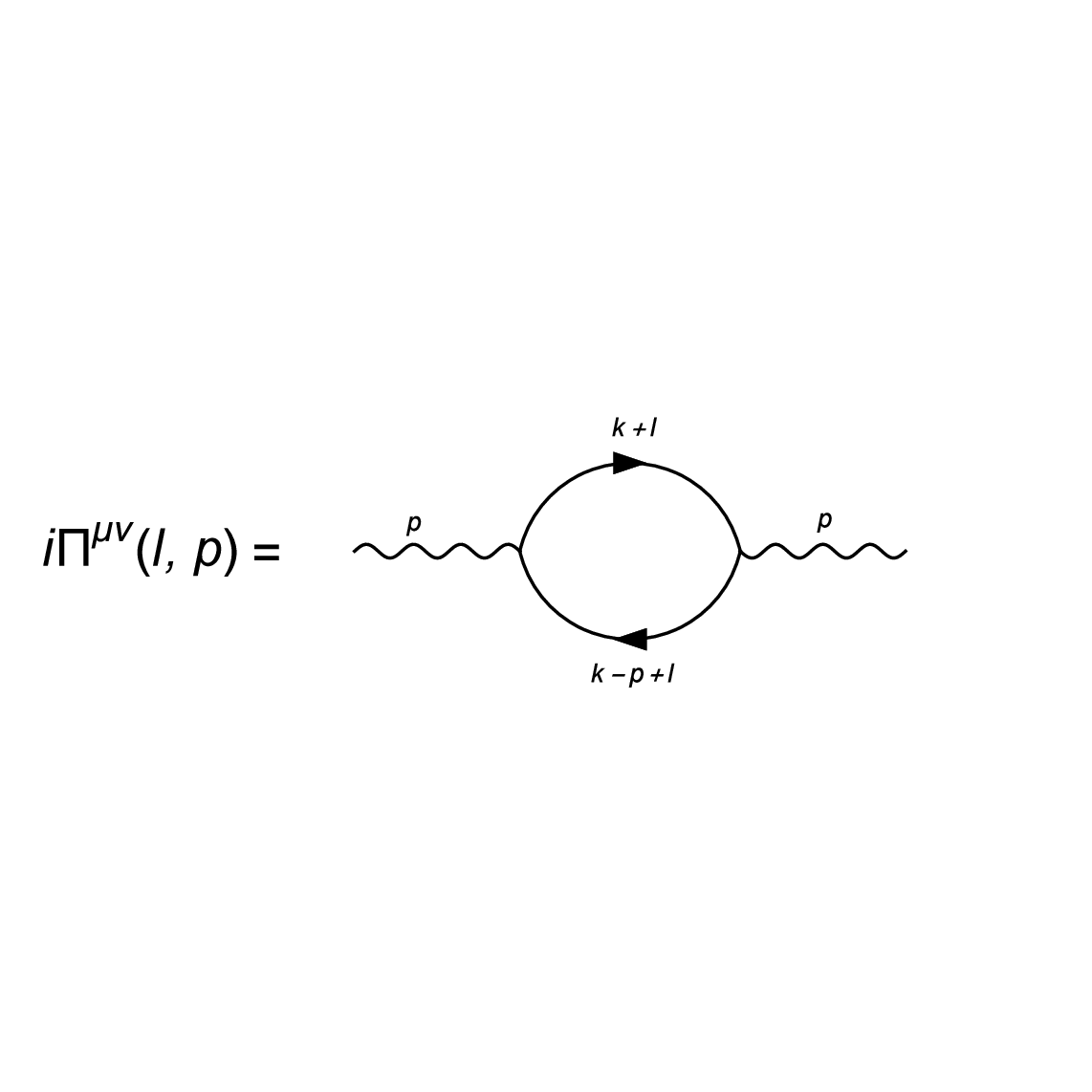}}
\caption{(a) One-loop 3-point Feynman diagram with the energy-momentum tensor as a vertex represented by a cross for a general momentum routing $l$. (b) Vacuum 
polarization tensor for a general momentum routing $l$.}
\label{fig4}
\end{figure}

Finally, $\Gamma^{\mu\nu}_{\mathcal{T}}(0,p)$ corresponds to the 3-point Feynman diagram $\Delta^{\mu\nu}(0,p,-p)\equiv \Delta^{\mu\nu}(p)$ shown in figure \ref{fig4}, 
where two of the vertices are the usual ones and the other is the trace of the energy-momentum tensor $\mathcal{T}^{\mu}_{\ \ \mu}=m\bar{\psi}\psi$,  
$\Gamma^{\mu\nu}(p)$ corresponds to the 2-point Feynman diagram $\Pi^{\mu\nu}(p,-p)\equiv \Pi^{\mu\nu}(p)$, which is the vacuum polarization tensor. In terms 
of these diagrams the Ward identity can be written as:
\be
\Delta^{\mu\nu}(p)=\left(2-p^{\lambda}\frac{\partial}{\partial p^{\lambda}}\right)\Pi^{\mu\nu}(p).
\label{eqIWD}
\ee

Our purpose in this section is to compute the scale anomalous term adding a generic momentum $l$ in the internal lines of the diagrams, as presented in
figure \ref{fig4}. So, the anomaly will be given by the violation of the equality (\ref{eqIWD}) for an arbitrary $l$: 
\be
\Delta^{\mu\nu}(l,p)=\left(2-p^{\lambda}\frac{\partial}{\partial p^{\lambda}}\right)\Pi^{\mu\nu}(l,p).
\label{eqIWD2}
\ee

The amplitudes of the Green functions involved in the relation (\ref{eqIWD2}) are given by

\begin{align}
&\Delta_{\mu\nu}(l,p)=-2\int \frac{d^4k}{(2\pi )^4} Tr[e \gamma_{\mu}S(k+l)e \gamma_{\nu}S(k+p+l)m S(k+p+l)],\nonumber\\
&i\Pi_{\mu\nu}(l,p)= \int \frac{d^4k}{(2\pi )^4} Tr[e \gamma_{\mu}S(k+l)e \gamma_{\nu} S(k+p+l)],
\label{Amps}
\end{align}
where $S(k)=\frac{i}{\slashed{k}-m}$ is the fermion propagator and $e$ is the electron charge.

We list all finite and regularized divergent integrals in the appendix. The result of the amplitude in eq. (\ref{Amps}) is given by
\bq
\Delta_{\mu\nu}(l,p)=\Delta_{\mu\nu}(p)=\frac{m^2 e^2}{2\pi^2}(g_{\mu\nu}Z_0+\iota_0 g_{\mu\nu}p^2-2\iota_1 p_{\mu}p_{\nu}-16\pi^2 i\upsilon_0 g_{
\mu\nu}),
\label{eqdelta}
\eq
where we see that this amplitude does not depend on the general routing $l$. This is expected since the linear divergent integral of the amplitude cancels.
In eqs. (\ref{Amps}), we see for the amplitude $\Delta^{\mu\nu}(l,p)$ that after multiplying both the numerator and denominator of the fermion propagator by $\slashed{k}+m$, the number of Dirac matrices inside the trace is odd for the linear divergent integral and therefore this trace is zero. So, the remaining divergent integrals are logarithmic divergent but all logarithmic divergent integrals are already momentum routing invariant. 

The other amplitude we need to compute to find out the scale anomaly was already presented in section \ref{sMRI} in eq. (\ref{eqpi2}). We also need to introduce the renormalization group scale $\lambda$ with the use of the scale relation below:
\be
I_{log}(m^2)=I_{log}(\lambda^2)+ b \ln \left(\frac{\lambda^2}{m^2}\right),
\label{eqSR}
\ee
where 
\begin{equation}
b\equiv \frac{i}{(4\pi)^2}.
\label{bdef}
\end{equation}

When the massless limit is taken, $\lambda$ is the only mass scale remaining. This is the part of the renormalization procedure where a renormalization scale is
introduced, as the $\mu$ scale in dimensional regularization. The $\lambda$ scale is intrinsic of the implicit regularization scheme. We notice that eq. 
(\ref{2.1}) splits an ultraviolet divergent integral in two infrared divergent ones in the massless limit. As a result, this scale $\lambda$ assures that
no infrared divergent integral remains in the final result as we shall see below.

As an example, let us consider the sum of the logarithmic divergent integral $I_{log}(m^2)$ with the ultraviolet finite integral $Z_0(p^2,m^2)$. A common 
appearance of these terms in the amplitude is $I_{log}(m^2)-bZ_0(p^2,m^2)$ and it can be rewritten using the scale relation (\ref{eqSR}):
\begin{align}
I_{log}(m^2)-bZ_0(p^2,m^2)&=I_{log}(\lambda^2)+ b \ln \left(\frac{\lambda^2}{m^2}\right)-b\int^1_0 dx \ln \left[\frac{m^2-p^2x(1-x)}{m^2}\right]=\nonumber\\
&=I_{log}(\lambda^2)-b\int^1_0 dx \ln \left[\frac{m^2-p^2x(1-x)}{\lambda^2}\right],
\end{align}
such that when we take the massless limit $m\rightarrow 0$ we find \[\lim_{m\rightarrow 0} (I_{log}(m^2)-bZ_0(p^2,m^2))=I_{log}(\lambda^2)-b\ln\left(-\frac
{p^2}{\lambda^2}\right)+
2b\] and the theory is infrared safe.

With all these pieces inserted in eq.(\ref{eqIWD2}) and taking the massless limit, we have
\begin{align}
\Delta^{\mu\nu}_r(l,p)-\left(2-p^{\lambda}\frac{\partial}{\partial p^{\lambda}}\right)\Pi^{\mu\nu}_r(l,p,\lambda)=& -4e^2 (\xi_0-2\upsilon_0 ) \Big(2 l^2 g^{\mu
\nu }+g^{\mu  \nu } l\cdot p+l^{\nu } p^{\mu }+l^{\mu } p^{\nu }+4 l^{\mu } l^{\nu }\Big)-8 e^2 \upsilon_2 g^{\mu  \nu } \nonumber\\
&+\frac{e^2}{6\pi^2}(p^2g^{\mu\nu}-p^{\mu}p^{\nu}),
\label{result}
\end{align}
where the index $r$ in the tensor stands for the renormalized amplitudes. It is interesting to notice that the divergent piece of eq. (\ref{eqIWD2})
besides being gauge invariant is scale invariant. On the other hand, the scale invariance of the finite piece is broken by a gauge invariant term. 

We see in eq. (\ref{result}) that in order to recover the correct result of the scale anomaly \cite{{Ellis}, {LoopReg}}, we must have $\upsilon_2=0$ and 
$\xi_0=2\upsilon_0$. This choice also leads to a result valid for any momentum routing $l$. This is the same condition required by gauge symmetry in section 
\ref{sMRI}. So, we can also conclude that gauge and momentum routing invariances imply in the correct result for the scale anomaly.

It is also interesting to notice that for massive theories the arbitrary surface term $\upsilon_0$ remains in eq. (\ref{eqdelta}) and it is not multiplying the
arbitrary momentum routing $l$ as in the examples of the previous sections or like the result of eq. (\ref{result}). In this case, this surface term can not be 
fixed by requiring consistency of scale symmetry breaking because we have already taken the massless limit in order to compute the anomalous term.

\section{The chiral anomaly in a non-minimal dimension-5 Lorentz-violating QED}
\label{sTLV}

As an application of the ideas presented in the previous sections, we compute the Ward identities of the triangle diagrams in a
Lorentz-violating QED in this section. The Standard Model Extension (SME) is a framework introduced in the late 90's to study possible 
violations of CPT and Lorentz symmetries \cite{{SME},{SME2}}. According to the spontaneous symmetry breaking (SSB) view, these symmetries are exact only 
at Planck scale in a string theory \cite{AlanSam} and as a result of this SSB there are several possible CPT and Lorentz-violating terms in 
the Standard Model Lagrangian at low energies. However, even if CPT and Lorentz symmetries are in fact exact in nature, the question 
is on what precision we can say they are. In this sense, there are numerous precision experiments performed for each sector of the 
SME. Precise tests with electrons \cite{electrons0}-\cite{electrons3} or the pure photon sectors \cite{photon1}-\cite{photon4} show how good 
are the CPT and the Lorentz symmetries in the QED sector of the SME.

The minimal SME includes only renormalizable operators while the non-minimal SME allows for non-renormalizable ones. In both
cases these operators bring consequences even for classical field theories like space-time anisotropy \cite{SME}, vacuum Cherenkov radiation 
\cite{{Cherenkov1},{Cherenkov2}}, vacuum birefringence \cite{{SME},{Carroll}}, gravitational waves \cite{Leila}-\cite{Petrov} or modifications in the 
polarization states of light \cite{SME}. In spectroscopy, both isotropic and anisotropic modes of CPT and Lorentz-violating operators can split the electronic 
energy levels in hydrogen and anti-hydrogen \cite{Neil}. It is interesting to know what are the consequences of breaking CPT or Lorentz symmetry at tree 
level. Not all coefficients for Lorentz violation that appear in the classical theory, prevail after one-loop corrections or a given operator at tree level 
can induce a different Lorentz-violating operator at the quantum level. For instance, the dimension-3  term $b^{\mu}\overline{\psi}\gamma_5\gamma_{\mu}\psi$ 
from the minimal fermion sector can radiatively induce the Chern-Simons-like term, which is a Lorentz-violating term of the minimal photon sector. The 
non-minimal term $(H_F^{(5)})^{\mu\nu\alpha\beta}\overline{\psi}\sigma_{\mu\nu}F_{\alpha\beta}\psi$ was studied in \cite{Manoel}, where the authors showed 
that this term induces radiatively the minimal term $(k_F)^{\mu\nu\alpha\beta}F_{\alpha\beta}F_{\mu\nu}$ of the photon sector of the SME. There was also a 
study of the chiral anomaly using the Fujikawa approach considering the term $b^{\mu}\overline{\psi}\gamma_5\gamma_{\mu}\psi$ from the fermion sector of the 
minimal SME and dimension-5 terms in the photon sector of the non-minimal SME\cite{AntAnomaly}.

Here we consider the dimension-5 operators of the fermion sector in the lagrangian $\mathcal{L}^{(5)}_{\psi F}$ of table I of reference \cite{Zonghao}. The 
terms with coefficients $(m_F^{(5)})^{\alpha\beta}$ and $(m_{5F}^{(5)})^{\alpha\beta}$  do not 
contribute to the chiral anomaly at first order because the number of Dirac matrices in the trace of the fermion loop is odd. One of 
the operators that produces non-trivial effects at one-loop is the dimension-5 CPT and Lorentz-violating non-minimal term
$-\frac{1}{2}(a^{(5)}_F)^{\mu\alpha\beta}\overline{\psi}\gamma_{\mu}F_{\alpha\beta}\psi$, where $(a^{(5)}_F)^{\mu\alpha\beta}$ is a
set of CPT and Lorentz-violating coefficients. It can be rewritten as 
$-(a^{(5)}_F)^{\mu\alpha\beta}\overline{\psi}\gamma_{\mu}\partial_{\alpha}A_{\beta}\psi$ due the anti-symmetry of the two last indices 
and this leads to the Feynman rule presented in figure \ref{figFR}.

\begin{figure}[!h]
\centering
\centering
\includegraphics[trim=0mm 160mm 35mm 0mm, scale=0.4]{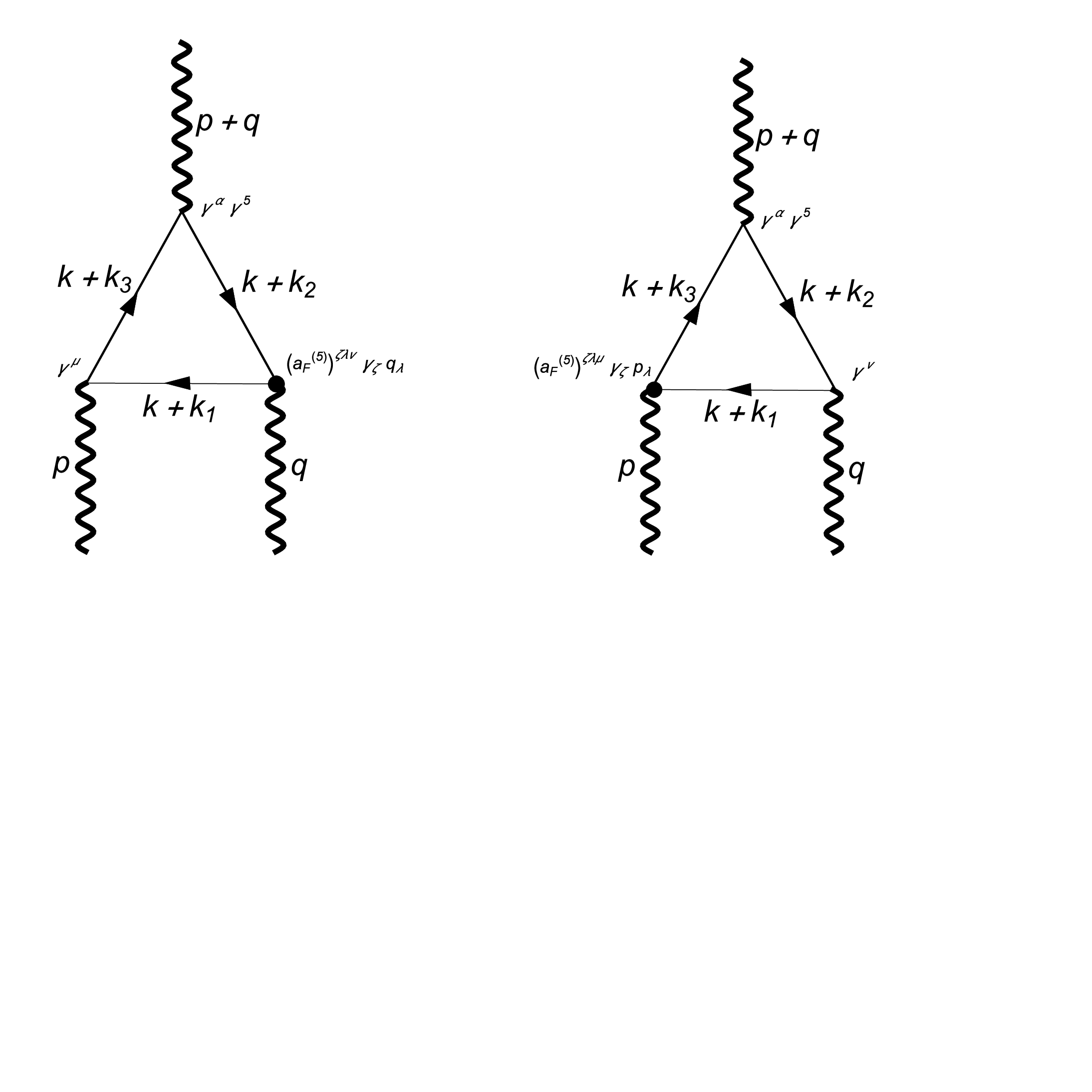}
\caption{Triangle diagrams in a QED with non-minimal Lorentz violation.}
\label{fig8.0}
\end{figure}

The triangle diagrams are depicted in figure \ref{fig8.0}. As in the usual case, this contribution must be summed with the crossed diagrams. Their corresponding amplitudes can be written as 

\begin{align}
T^{LV}_{\mu\nu\alpha}&=-i\int_k Tr\left[\gamma_{\mu}\frac{i}{\slashed{k}+\slashed{k}_1-m}(a_F^{(5)})_{\zeta\lambda\nu}\gamma^{\zeta}q^{\lambda}\frac{i}{\slashed{k}+\slashed{k}_2-m}\gamma_{\alpha}\gamma^5 \frac{i}{\slashed{k}+\slashed{k}_3-m}\right]+\nonumber\\
&-i\int_k Tr\left[(a_F^{(5)})^{\zeta\lambda\mu}\gamma_{\zeta}p_{\lambda}\gamma_{\mu}\frac{i}{\slashed{k}+\slashed{k}_1-m}\gamma_{\nu}\frac{i}{\slashed{k}+\slashed{k}_2-m}\gamma_{\alpha}\gamma^5 \frac{i}{\slashed{k}+\slashed{k}_3-m}\right]
+(\mu \leftrightarrow \nu, p\leftrightarrow q).
\label{AVVLV}
\end{align}

The parameterization of the arbitrary routing $k_i$ is according to the eqs. (\ref{routing2}). Now, if we compute the Ward identities we find out as expected that gauge symmetry is broken by the arbitrary surface term $\upsilon_0$ as presented below
\begin{align}
&p_{\mu}T^{LV\mu\nu\alpha}=-8i\beta\upsilon_0 (a^{(5)}_{F})^{\ \ \ q\nu}_{\zeta}\epsilon^{q\alpha\zeta p},\nonumber\\
&q_{\nu}T^{LV\mu\nu\alpha}=8i(1-\alpha )\upsilon_0(a^{(5)}_{F})^{\ \ \ p\mu}_{\zeta}\epsilon^{q\alpha\zeta p},\nonumber\\
(p_{\alpha}+q_{\alpha})T^{LV\mu\nu\alpha}&=8i(\alpha-\beta-1 )\upsilon_0((a^{(5)}_{F})^{\ \ \ q\nu}_{\zeta}\epsilon^{q\zeta\mu p}-(a^{(5)}_{F})^{\ p\mu}_{\zeta}\epsilon^{q\zeta\nu p})\nonumber\\
&-\frac{1}{2\pi^2}((a^{(5)}_{F})^{\ \ \ q\nu}_{\zeta}\epsilon^{q\zeta\mu p}+(a^{(5)}_{F})^{\ \ \ p\mu}_{\zeta}\epsilon^{p\zeta\nu q})(1+2m^2\xi_{00}(p,q)),
\label{IWsLV}
\end{align}
where $(a^{(5)}_F)^{\alpha p\mu}\equiv (a^{(5)}_F)^{\alpha \lambda\mu}p_{\lambda}$, $\epsilon^{\alpha\lambda\nu p}\equiv\epsilon^{\alpha\lambda\nu 
\sigma}p_{\sigma}$ and $2m^2\xi_{00}(p,q)$ is the contribution of the usual vector-vector-pseudoscalar diagram that vanishes in the massless limit. The
computation of the finite piece of these Ward identities is detailed in the appendix.

In this case, there is no particle process such as the pion decay where the breaking of chiral symmetry provided by Lorentz violation at tree level is measured.
So, the ambiguity in the momentum routing can not be solved by means of an experiment. However, since we showed in section \ref{ABJ} that momentum routing 
invariance is not broken in the usual chiral anomaly, we can resort to momentum routing invariance in eqs. (\ref{IWsLV}). We see that the results of the Ward 
identities are momentum routing invariant if the surface term $\upsilon_0$ is null. As a consequence, we have gauge symmetry and a chiral anomaly induced by 
CPT and Lorentz violation at the quantum level. Notice that the dimension-5 CPT and Lorentz-violating operator is both gauge and chiral invariant at the tree 
level but only gauge symmetry holds at the quantum one. It is noteworthy that some choices of momentum routing such as $\alpha=0$ and
$\beta=-1$ is compatible with the Ward identities of the Lorentz invariant situation measured by experiments, {\it i. e.} gauge symmetry holding with no chiral 
symmetry, as we can see from eqs. (\ref{WI22}). Nevertheless, this choice of momentum routing breaks both gauge and chiral symmetry in the Lorentz-violating 
situation as we can see in the Ward identities (\ref{IWsLV}). We can instead alternatively choose a null value of the surface term so that the Ward identities
of both cases are compatible with momentum routing invariance. In this case, we would have in both situations gauge symmetry and the breaking of chiral 
symmetry.

\section{Conclusions}
\label{sC}

In this work, we show that the choice of the internal momentum route in order to compute anomalies break the momentum routing 
invariance. However, this does not necessarily imply that momentum routing invariance of the Feynman diagrams is violated because the 
anomalies are physical. It is possible to compute the chiral and the scale anomalies in a momentum routing invariant way. So, even if 
these anomalies are present, they are still compatible with a general momentum routing. Because of the gauge and momentum routing 
invariance relation, we see that the latter invariance is violated if the former is as well for some specific diagrams. This conclusion is 
supported by implicit regularization since the conditions that assure gauge invariance are the same that assure the momentum routing 
one. Nevertheless, the results here presented does not invalidate, of course, previous computations of the chiral anomaly since in 
sections \ref{CAZ} and \ref{ABJ} we show that a gauge invariant result is also compatible with a specific momentum routing choice in 
the case of the chiral anomaly.

In summary, the contribution made by this paper resides in showing that anomalies does not depend on the momentum routing. It is possible to get the 
well known result of the scale and the chiral anomalies for a general momentum routing. Besides, this idea is
important specially for theories that are not yet experimentally accessible, like supersymmetric theories or frameworks with Lorentz
and CPT violation. In these cases, it would not be possible to solve the theoretical ambiguity of momentum routing by means of an
experiment. On the other hand, the ambiguity of the chiral anomaly due the momentum routing choice can be solved because of the 
measurement of the pion decay. In theories where the measurement of the anomaly is not available, one can resort to momentum routing 
invariance since it is shown in this paper that there is no evidence of the momentum routing invariance breaking. Based on this, we apply the idea
to a QED with CPT and Lorentz violation in section \ref{sTLV}. We see that if momentum routing invariance is assumed, gauge symmetry holds and
chiral symmetry is broken as in the Lorentz invariant case. However, there are momentum routing choices for diagrams of the Lorentz invariant case that assure
gauge symmetry with chiral symmetry breaking but break both symmetries in the Lorentz-violating situation. So, the momentum routing invariance requirement 
makes the new scenario compatible with the usual one.

As a prospect, it would be also interesting to perform the same analysis for gravitational anomalies such as the conformal anomaly in 
(3+1)-dimensions \cite{{Gazzola}, {Shapiro}} or the Einstein and Weyl anomalies in (1+1)-dimensions \cite{{BertlmannPLB},{Leo}}. It 
would be interesting to check for these gravitational theories if there is any relation between the momentum routing invariance of the 
Feynman diagrams and diffeomorphisms.

\section*{Appendix} 
\label{A}

\subsection*{Feynman rules}

\begin{figure}[!h]
\centering
\centering
\includegraphics[trim=0mm 125mm 0mm 0mm, scale=0.4]{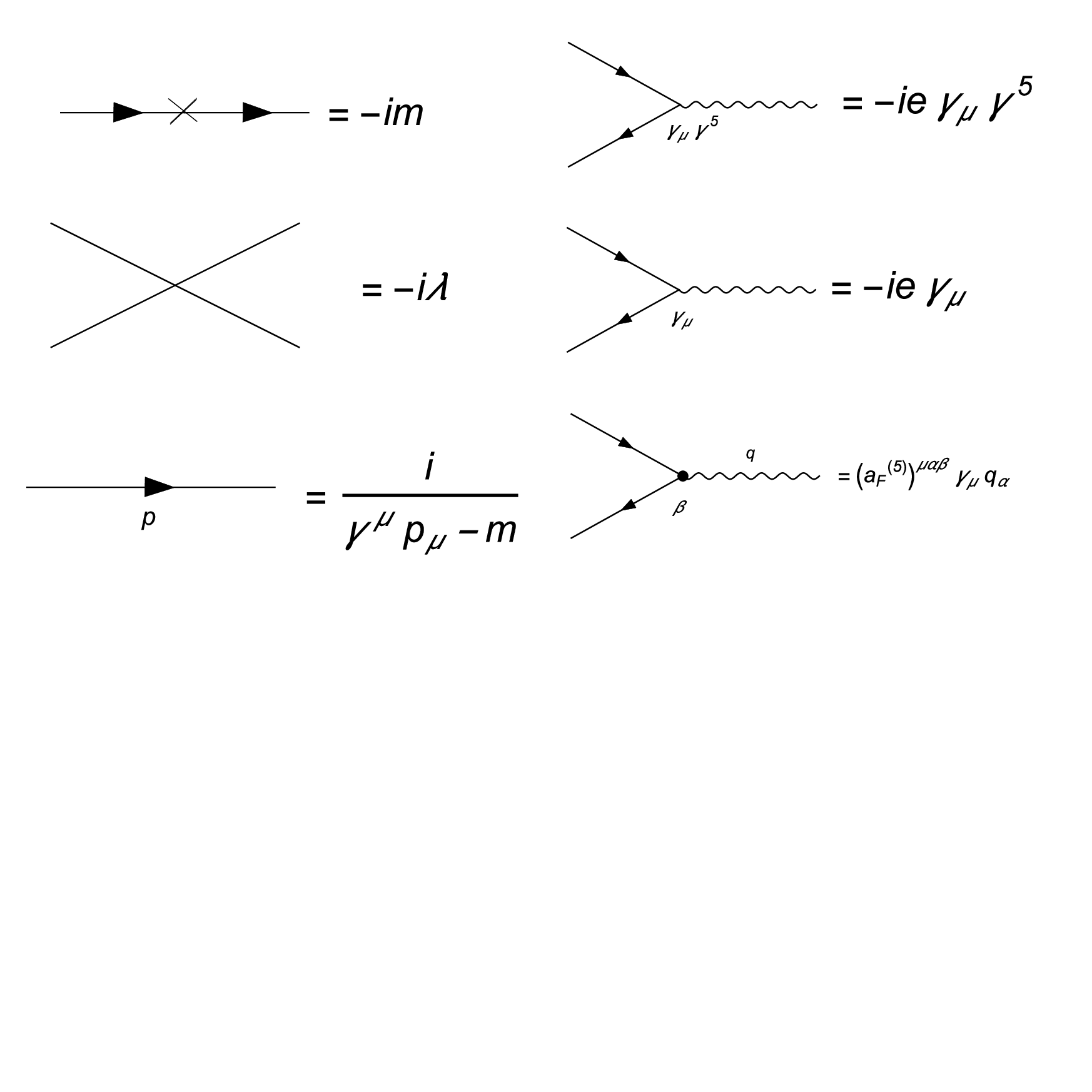}
\caption{Feynman rules used to build the amplitudes of sections \ref{sMRI}, \ref{CAZ}, \ref{ABJ}, \ref{sSA} and
\ref{sTLV}. The waved, solid and solid with arrow lines stand for photon, scalar and fermion, respectively. The dot means a Lorentz-violating interaction 
vertex.}
\label{figFR}
\end{figure}

The set of all Feynman rules needed to build the amplitudes discussed in this manuscript is listed in figure \ref{figFR}.

\subsection*{Finite part $T^{finite}_{\mu\nu\alpha}$ and $T^{LV\ finite}_{\mu\nu\alpha}$ of the triangle diagrams }

We perform the computation of the finite part of the triangle diagram, $T^{finite}_{\mu\nu\alpha}$. Since it does not
depend on the routing, we can choose $k_1=0$, $k_2=q$ e $k_3=-p$ and we have:

\begin{align}
T_{\mu\nu\alpha}=&-i\int_k Tr\left[\gamma_{\mu}\frac{i}{\slashed{k}-m}\gamma_{\nu}\frac{i}{\slashed{k}+\slashed
{q}-m}\gamma_{\alpha}\gamma^5 \frac{i}{\slashed{k}-\slashed{p}-m}\right]+(\mu \leftrightarrow \nu, p\leftrightarrow q)=\nonumber\\
&= -8i\upsilon_0 \epsilon_{\mu \nu \alpha \beta}(q-p)^{\beta}+T^{finite}_{\mu\nu\alpha},
\end{align}

After taking the trace and regularizing we find out the finite part of the amplitude. We list the results of the integrals in the final
part of this section. The result is

\begin{align}
T^{finite}_{\mu\nu\alpha}=& 4i b\{\epsilon_{\alpha\mu\nu\lambda}q^{\lambda}(p^2\xi_{01}(p,q)-q^2\xi_{10}(p,q))+\epsilon_{
\alpha\mu\nu\lambda}q^{\lambda}(1+2m^2\xi_{00}(p,q))+
\nonumber\\ & +4\epsilon_{\alpha\nu\lambda\tau}p^{\lambda}q^{\tau}[(\xi_{01}(p,q)-\xi_{02}(p,q))p_{\mu}+\xi_{11}(p,q)q_{\mu}]+
(\mu \leftrightarrow \nu, p\leftrightarrow q)\},
\label{Tfinito}
\end{align}
where the functions $\xi_{nm}(p,q)$ are defined as
\be
\xi_{nm}(p,q)=\int^1_0 dz\int^{1-z}_0 dy \frac{z^n y^m}{Q(y,z)},\\
\ee
with
\be
Q(y,z)=[p^2 y(1-y)+q^2 z(1-z)+2(p\cdot q)yz-m^2]
\ee
and those functions have the property $\xi_{nm}(p,q)=\xi_{mn}(q,p)$.

The $\xi_{nm}$ functions obey the following relations which we have already used in the derivation of eq. (\ref{Tfinito})
\bq
&q^2 \xi_{11}(p,q)-(p\cdot q)\xi_{02}(p,q)=\frac{1}{2}\left[ -\frac{1}{2}Z_0((p+q)^2,m^2)+\frac{1}{2}Z_0(p^2,m^2)+q^2 \xi_{01}(p,q)
\right],\label{f1}\\
&p^2 \xi_{11}(p,q)-(p\cdot q)\xi_{20}(p,q)=\frac{1}{2}\left[ -\frac{1}{2}Z_0((p+q)^2,m^2)+\frac{1}{2}Z_0(q^2,m^2)+p^2 \xi_{10}(p,q)
\right],\label{f2}\\
&q^2 \xi_{10}(p,q)-(p\cdot q)\xi_{01}(p,q)=\frac{1}{2}[ -Z_0((p+q)^2,m^2)+Z_0(p^2,m^2)+q^2 \xi_{00}(p,q)],\label{f3}\\
&p^2 \xi_{01}(p,q)-(p\cdot q)\xi_{10}(p,q)=\frac{1}{2}[ -Z_0((p+q)^2,m^2)+Z_0(q^2,m^2)+p^2 \xi_{00}(p,q)],\label{f4}\\
&q^2 \xi_{20}(p,q)-(p\cdot q)\xi_{11}(p,q)=\frac{1}{2}\left[-\left(\frac{1}{2}+m^2\xi_{00}(p,q)\right)+\frac{1}{2}p^2\xi_{01}(p,q)+\frac
{3}{2}q^2\xi_{10}(p,q)\right],\label{f5}\\
&p^2 \xi_{02}(p,q)-(p\cdot q)\xi_{11}(p,q)=\frac{1}{2}\left[-\left(\frac{1}{2}+m^2\xi_{00}(p,q)\right)+\frac{1}{2}q^2\xi_{10}(p,q)+\frac
{3}{2}p^2\xi_{01}(p,q)\right],\label{f6}
\eq
where $Z_k(p^2,m^2)$ is defined in eqs. (\ref{finiteint1}).

The derivation of the relations (\ref{f1})-(\ref{f6}) can be simply achieved by integration by parts. There is a whole review \cite
{Orimar2} about these integrals and other integrals with integrands of larger denominators that appear in Feynman diagrams with more external legs.

In the scenario with CPT and Lorentz violation, the computation of the finite part of the triangle diagrams, $T^{LV\ finite}_{\mu\nu\alpha}$,
is simpler to compute if the external momenta are contracted with the tensor of the amplitude. Here, we present an example of a gauge Ward identity as 
presented below

\begin{align}
p^{\mu}T^{LV}_{\mu\nu\alpha}=&-i\int_k Tr\left[\slashed{p}\frac{i}{\slashed{k}+\slashed{k}_1-m}(a_F^{(5)})_{\zeta\lambda\nu}\gamma^{\zeta}q^{\lambda}\frac{i}{\slashed{k}+\slashed{k}_2-m}\gamma_{\alpha}\gamma^5 \frac{i}{\slashed{k}+\slashed{k}_3-m}\right]+\nonumber\\
&-i\int_k Tr\left[(a_F^{(5)})^{\zeta\lambda\mu}\gamma_{\zeta}p_{\lambda}\slashed{p}\frac{i}{\slashed{k}+\slashed{k}_1-m}\gamma_{\nu}\frac{i}{\slashed{k}+\slashed{k}_2-m}\gamma_{\alpha}\gamma^5 \frac{i}{\slashed{k}+\slashed{k}_3-m}\right]
+(\mu \leftrightarrow \nu, p\leftrightarrow q)=\nonumber\\
&= 8i\beta (a^{(5)}_F)^{\zeta q}_{\ \ \ \nu} \epsilon _{\alpha  \zeta  p q}+T^{LV\ finite}_{\mu\nu\alpha},
\end{align}
where $(a^{(5)}_F)^{\zeta q\nu}\equiv(a^{(5)}_F)^{\zeta \lambda\nu}q_{\lambda}$ and $\epsilon _{\alpha  \zeta  \beta q}\equiv 
\epsilon _{\alpha\zeta\beta\lambda}q_{\lambda}$ .

Since the finite part does not depend on the routing, we can choose $k_1=0$, $k_2=q$ e $k_3=-p$ and we have the following result after taking the traces and 
using the results of the regularized integrals

\begin{align}
&p^{\mu}T^{LV\ finite}_{\mu\nu\alpha}=-4i b\Big\{-(a^{(5)}_F)^{\zeta q}_{\ \ \ \nu}\epsilon _{\alpha  \zeta  p q}p^2 \xi_{00}(p,q)
+2 (a^{(5)}_F)^{\zeta q}_{\ \ \ \nu}\epsilon_{\alpha \zeta p q}p^2 \xi _{01}(p,q)+ 2(a^{(5)}_F)^{\zeta q}_{\ \ \ \nu}
\epsilon_{q\alpha  \zeta p} p\cdot q \xi_{10}(p,q)\nonumber\\
&+\frac{1}{2}(a^{(5)}_F)^{\zeta q}_{\ \ \ \nu}\epsilon_{q\alpha  \zeta  p}[Z_0(q^2,m^2)-Z_0((p+q)^2, m^2)]+\frac{1}{2} (a^{(5)}_F)^{\zeta q}_{\ \ \ \nu}
\epsilon_{q\alpha  \zeta  p}[Z_0(p^2,m^2)-Z_0((p+q)^2, m^2)]\nonumber\\
&-\frac{1}{2}(a^{(5)}_F)^{\zeta q}_{\ \ \ \nu}\epsilon_{q\alpha  \zeta  p}Z_0(p^2,m^2)+\frac{1}{2}(a^{(5)}_F)^{\zeta q}_{\ \ \ \nu}\epsilon_{q\alpha  
\zeta p}Z_0(q^2,m^2) \Big\},
\end{align}
where eqs. (\ref{f1})-(\ref{f6}) were used to simplify the result. The main idea is to use relations (\ref{f1})-(\ref{f2}) and (\ref{f5})-(\ref{f6}) to reduce 
integrals $\xi_{20}(p,q)$, $\xi_{02}(p,q)$ and $\xi_{11}(p,q)$ into integrals $\xi_{10}(p,q)$ and $\xi_{01}(p,q)$. Then the relations (\ref{f3})-(\ref{f4}) 
are used to reduce everything in terms of $\xi_{00}(p,q)$ and $Z_n$'s. After this procedure, we find as expected that the finite part is gauge invariant: 
\begin{equation}
p^{\mu}T^{LV\ finite}_{\mu\nu\alpha}=0.
\end{equation}

The same idea is applied in the computation of the other two Ward identities.

\subsection*{Result of the integrals of sections \ref{sMRI} and \ref{sSA} }

The result of all finite and regularized divergent integrals from sections \ref{sMRI} and \ref{sSA} is listed below:

\begin{align}
&\int_k \frac{k^{\mu}}{(k^2-m^2)^2[(k-p)^2-m^2]}=-b p^{\mu} \iota_1 ,\\
&\int_k \frac{1}{[(k+p)^2-m^2]^2}=I_{log}(m^2),\\
&\int_k \frac{1}{[(k+p)^2-m^2]}=I_{quad}(m^2)-p^2\upsilon_0,\\
&\int_k \frac{1}{(k^2-m^2)[(k+p)^2-m^2]^2}=\int_k \frac{1}{(k^2-m^2)^2[(k+p)^2-m^2]}=-b \iota_0 ,\\
&\int_k \frac{k^{\mu}}{(k^2-m^2)[(k+p)^2-m^2]^2}=-b p^{\mu} \iota_1+ b p^{\mu}\iota_0 ,\\
&\int_k \frac{k^{\mu}k^{\nu}}{(k^2-m^2)[(k+p)^2-m^2]^2}=\frac{1}{4}g_{\mu \nu}(I_{log}(m^2)-\upsilon_0)-\frac{1}{2} g_{\mu \nu} b(Z_0-Z_1)-b(\iota _0-2 \iota _1+\iota _2)p^{\mu} p^{\nu}, \\
&\int_k \frac{k^{\mu}}{[(k+l)^2-m^2][(k+p+l)^2-m^2]}=  -\left( l^{\alpha}+\frac{p^{\alpha }}{2}\right)( I_{log}(m^2)-\upsilon_0 )+\frac{b}{2}p^{\mu} Z_0, \\
\end{align}
\begin{align}
&\int_k \frac{k^{\mu}k^{\nu}}{[(k+l)^2-m^2][(k+p+l)^2-m^2]}=\frac{1}{2}g^{\mu  \nu }(I_{quad}(m^2)-\upsilon_2) -\frac{1}{4} g^{\mu  
\nu }(I_{log}(m^2)-\upsilon_0 ) \left( 2 l^2+2 l\cdot p+p^2\right)+ \nonumber\\
&+\frac{1}{6} (I_{log}(m^2)-\xi_0 ) \left(g^{\mu  \nu } \left(3 l^2+3 l\cdot p+p^2\right)+3 l^{\nu } p^{\mu }+3 l^{\mu } p^{\nu
}+6 l^{\mu } l^{\nu }+2 p^{\mu } p^{\nu }\right)+b\left(\frac{1}{3}p^{\mu}p^{\nu}+b\frac{1}{12}p^{2}g^{\mu \nu} \right) Z_0- \nonumber\\
&-\frac{1}{4}g^{\mu\nu}p^2 \upsilon_0-\frac{1}{3}b(p^2g^{\mu\nu}-p^{\mu}p^{\nu})\left(\frac{m^2}{p^2}Z_0 +\frac{1}{6} \right), \nonumber\\
\end{align}
  
where $Z_n(p^2,m^2)$, $\iota_n$ and $D(x)$ are defined in eqs. (\ref{finiteint1})-(\ref{finiteint3}) and $b$ is defined in eq. (\ref{bdef}).

\subsection*{Result of the integrals of section \ref{ABJ}}

\begin{equation}
\int_k \frac{1}{(k^2-m^2)[(k-p)^2-m^2][(k+q)^2-m^2]}= b \xi_{00}(p,q),
\end{equation}
\begin{equation}
\int_k \frac{k^{\alpha}}{(k^2-m^2)[(k-p)^2-m^2][(k+q)^2-m^2]}= b (p^{\alpha}\xi_{01}(p,q)-q^{\alpha}\xi_{10}(p,q)),
\end{equation}
\begin{align}
\centering
\int_k \frac{k^{2}}{(k^2-m^2)[(k-p)^2-m^2][(k+q)^2-m^2]}=& I_{log}(m^2)-b Z_0(q^2,m^2)+ b (m^2-p^2)\xi_{00}(p,q)+\nonumber\\
&+2b(p^2\xi_{01}(p,q)-(p\cdot q) \xi_{10}(p,q)),
\end{align}
\begin{align}
&\int_k \frac{k^{\alpha}k^{\beta}}{(k^2-m^2)[(k-p)^2-m^2][(k+q)^2-m^2]}= \frac{1}{4}g^{\alpha\beta}(I_{log}(m^2)-\upsilon_0)-\frac
{1}{4}b g^{\alpha\beta} Z_0(q^2,m^2)-\nonumber\\&-b\Big[\frac{1}{2}g^{\alpha\beta}p^2(\xi_{00}(p,q)-3\xi_{01}(p,q)-\xi_{10}(p,q)+2\xi_
{02}(p,q)+2\xi_{11}(p,q))-\xi_{02}(p,q)p^{\alpha}p^{\beta}+\nonumber\\&+\xi_{11}(p,q)q^{\alpha}p^{\beta}+\xi_{11}(p,q)p^{\alpha}q^{\beta}
-\xi_{20}(p,q)q^{\alpha}q^{\beta}+(\xi_{10}(p,q)-\xi_{11}(p,q)-\xi_{20}(p,q))g^{\alpha\beta}(p\cdot q)\Big],
\end{align}
\begin{align}
&\int_k \frac{k^{\alpha}k^{2}}{(k^2-m^2)[(k-p)^2-m^2][(k+q)^2-m^2]}=\frac{1}{2}(p^{\alpha}-q^{\alpha})(I_{log}(m^2)-\upsilon_0)+
\frac{1}{2}b(q^{\alpha}Z_0(q^2,m^2)-\nonumber\\ & -p^{\alpha}Z_0(p^2,m^2))+ b(m^2- q^2)(p^{\alpha}\xi_{01}(p,q)-q^{\alpha}\xi_{10}(p,q))+
b[q^{\alpha}p^2(\xi_{00}(p,q)-3\xi_{01}(p,q)-\xi_{10}(p,q)+\nonumber\\ &+2\xi_{02}(p,q)+2\xi_{11}(p,q))-2(p\cdot q)p^{\alpha}\xi_{02}(p,q)
+2 q^2 p^{\alpha}\xi_{11}(p,q)+2(p\cdot q)q^{\alpha}(\xi_{10}(p,q)-\nonumber\\&-\xi_{20}(p,q))-2 q^2 q^{\alpha}\xi_{20}(p,q))],
\end{align}
where  $\int_k\equiv \int^{\Lambda} \frac{d^4 k}{(2\pi)^4}$ and $b$ is defined in eq. (\ref{bdef}).







\begin{thebibliography}{99}
\bibitem{Costa} M. Costa, H. Panagopoulos, Phys. Rev. D 96 (2017) 034507.
\bibitem{Cynolter0} G. Cynolter, E. Lendvai, Mod. Phys.Lett. A 29 (2014) 1450024.
\bibitem{Jackiw} J.S. Bell, R. Jackiw, Nuovo Cim. A 60 (1969) 47-61.
\bibitem{Adler} S.L. Adler, Phys. Rev. 177 (1969) 2426.
\bibitem{Bardeen} W.A. Bardeen, Phys. Rev. 184 (1969) 1848.
\bibitem{Ellis} M.S. Chanowitz, J. Ellis, Phys. Rev. D  7 (1973) 2490.
\bibitem{Ahmad} A. Ahmad, G. Varma K., G. Sharma, J. Phys. Condens. Matter 37 (2025) 043001.
\bibitem{Claudio} C. Corian\`o, M. Cret\`i, S. Lionetti, R. Tommasi., Phys. Rev. D 110 (2024) 025014.
\bibitem{Hao} H. Dang, Z. Xing, M.A. Sultan, K. Raya, L. Chang, Phys. Rev. D 108 (2023) 054031.
\bibitem{glueball} A. Ballon-Bayona, H. Boschi-Filho, L.A.H. Mamani, A.S. Miranda, V.T. Zanchin, Phys. Rev. D 97 (2018) 046001.
\bibitem{Xing} Z. Xing, H. Dang, M.A. Sultan, K. Raya, L. Chang, Phys. Rev. D 109 (2024) 054028.
\bibitem{Nielsen} H.B. Nielsen, M. Ninomiya, Phys. Lett. B 130 (1983) 389.
\bibitem{PRB} V.A. Zyuzin, Phys. Rev. B 95 (2017) 245128.
\bibitem{Science} J. Xiong {\it et al.}, Science 350 (2015) 413.
\bibitem{Nature1} M. Hirschberger {\it et al.}, Nature Materials 15 (2016) 1161.
\bibitem{Nature2} Cheng-Long Zhang {\it et al.}, Nature Communications 7 (2016) 10735. 
\bibitem{Review} N.P. Armitage, E.J. Mele, A. Vishwanath, Rev. of Mod. Phys.  90 (2018) 015001.
\bibitem{Prereg} V. Elias, G. McKeon, S.B. Phillips, R.B. Mann, Phys. Lett. B 133 (1983) 83.
\bibitem{McKeon} A.M. Chowdhury, V. Elias, D.G.C. McKeon, R.B. Mann, Phys. Rev. D 34 (1986) 619.
\bibitem{IR} O.A. Battistel, A.L. Mota, M.C. Nemes, Mod. Phys. Lett. A 13 (1998) 1597.
\bibitem{DR} G. 't Hooft, M. Veltman, Nucl. Phys. B 44 (1972) 189.
\bibitem{Bollini} C.G. Bollini, J.J. Giambiagi, Nuovo Cimento 12 (1972) 20.
\bibitem{Siegel} W. Siegel, Phys. Lett. B 84 (1979) 193.
\bibitem{Siegel2} W. Siegel, Phys. Lett. B 94 (1980) 37.
\bibitem{Orimar} G. Dallabona, P.G. de Oliveira, O.A. Battistel, J. Phys. G 51 (2024) 095004.
\bibitem{Ricardo} R.J.C. Rosado, A. Cherchiglia, M. Sampaio, B. Hiller, Eur. Phys. J. C 85 (2025) 41.
\bibitem{Ricardo2} R.J.C. Rosado, A. Cherchiglia, M. Sampaio, B. Hiller,  Acta Phys. Polon. Supp. 17 (2024) 6-A15.
\bibitem{Adriano3} A.M. Burque, A.L. Cherchiglia, M. P\'erez-Victoria, JHEP 08 (2018) 109.
\bibitem{JackiwFU} R. Jackiw , Int. J. Mod. Phys. B 14 (2000) 2011.
\bibitem{Bruno} B.Z. Felippe, A.P.B. Scarpelli, A. R. Vieira, J.C.C. Felipe, Eur. Phys. J. C 82 (2022) 583.
\bibitem{Pugh} R.E. Pugh, Can. J. of Phys. 47 (1969) 1263.
\bibitem{Adriano0} A.L. Cherchiglia, M. Sampaio, M.C. Nemes, Int. J. Mod. Phys. A 26 (2011) 2591-2635. 
\bibitem{Peskin} M.E. Peskin, D. Schroeder, An Introduction to Quantum Field Theory, Addison-Wesley, 1995.
\bibitem{MRICGT} A.C.D. Viglioni, A.L. Cherchiglia, A.R. Vieira, B. Hiller, M. Sampaio, Phys. Rev. D 94 (2016) 065023.
\bibitem{Adriano} L.C. Ferreira, A.L. Cherchiglia, B. Hiller, M. Sampaio, M.C. Nemes, Phys. Rev.D 86 (2012) 025016. 
\bibitem{MRILV} A.R. Vieira, A.L. Cherchiglia, M. Sampaio, Phys. Rev. D 93 (2016) 025029.
\bibitem{Zee2} A. Zee, Quantum Field Theory in a Nutshell, Cambridge University Press, 2003.
\bibitem{Bertlmann} R.A. Bertlmann, Anomalies in Quantum Field Theory, Oxford University Press, 1996.
\bibitem{Cynolter} G. Cynolter and E. Lendvai, Mod. Phys. Lett. A 26 (2011) 1537.
\bibitem{Luciana} J.F. Thuorst, L. Ebani, T.J. Girardi, Annals of Phys. 468 (2024) 169725. 
\bibitem{Elias} V. Elias, G. McKeon, R. B. Mann, Nucl. Phys. B 229 (1983) 487.
\bibitem{YU} H.-L. Yu, W. B. Yeung, Phys. Rev. D 35 (1987) 3955.
\bibitem{Chen} L. Chen, JHEP 2023 (2023) 30.
\bibitem{Tsai} Er-Cheng Tsai, Phys. Rev. D 83 (2011) 025020.
\bibitem{Tsai2} Er-Cheng Tsai, Phys. Rev. D 83 (2011) 065011.
\bibitem{Fujikawa} K. Fujikawa, Phys. Rev. Lett. 42 (1979) 1195.
\bibitem{Fujikawa2} K. Fujikawa, Phys. Rev. D 21 (1980) 2848.
\bibitem{Zee} B. Zumino, W. Yong-Shi, A. Zee, Nucl. Phys. B 239 (1984) 477.
\bibitem{Perez} F. del Aguila, M. Perez-Victoria, Acta Phys. Pol. B 29 (1998) 2857.
\bibitem{Wu} Y. L. Ma, Y. L. Wu, Int. J. Mod. Phys. A 21 (2006) 6383.
\bibitem{tHooft} G. t'Hooft, Phys. Rev. Lett. 37 (1976) 8.
\bibitem{Adriano2} A.L. Cherchiglia, Nucl. Phys. B 987 (2023) 116104. 
\bibitem{Callan} C.G. Callan, Phys. Rev. D 2 (1970) 1541. 
\bibitem{LoopReg} J.-W. Cui, Y.-L. Ma, Y.-L. Wu, Phys. Rev. D 84 (2011) 025020.
\bibitem{Collins} J.C. Collins, A. Duncan, S.D. Joglekar, Phys. Rev. D 16 (1977) 438.
\bibitem{SME} D. Colladay, V.A. Kosteleck\'y, Phys. Rev. D 58 (1998) 116002.
\bibitem{SME2}  D. Colladay, V.A. Kosteleck\'y, Phys. Rev. D 55 (1997) 6760.
\bibitem{AlanSam} V.A. Kosteleck\'y, S. Samuel, Phys. Rev. D 39 (1989) 683.
\bibitem{electrons0} M.J. Borchert et al., Nature 601 (2022) 53-57.
\bibitem{electrons1} Y. Ding, M.F. Rawnak, Phys. Rev. D 102 (2020) 056009.
\bibitem{electrons2} T. Pruttivarasin et al., Nature 517 (2015) 592.
\bibitem{electrons3} V.A. Kosteleck\'y, A.J. Vargas, Phys.\ Rev.\ D 92 (2015) 056002.
\bibitem{photon1} J.-Q. Xia, H. Li, X. Zhang, Phys.\ Lett.\ B 687 (2010) 129.
\bibitem{photon2} M.L. Brown et al., Astrophys.\ J.\  705 (2009) 978.
\bibitem{photon3} V.A. Kosteleck\'y, M. Mewes, Phys.\ Rev.\ D 80 (2009) 015020 .
\bibitem{photon4} M. Mewes, Phys.\ Rev.\ D 78 (2008) 096008.
\bibitem{Cherenkov1} M. Schreck, Phys. Rev. D 96 (2017) 095026.
\bibitem{Cherenkov2} M. Schreck, Symmetry 10 (2018) 424.
\bibitem{Carroll} S.M. Carroll, G.B. Field, R. Jackiw, Phys. Rev. D 41 (1990) 1231.
\bibitem{Leila} L. Haegel et al., Phys. Rev. D 107 (2023) 6.
\bibitem{Kellie} K. O'Neal-Ault, Q.G. Bailey, T. Dumerchat, L. Haegel, J. Tasson, Universe 7 (2021) 380.
\bibitem{Petrov} J.R. Nascimento, A.Yu. Petrov, A.R. Vieira, Galaxies 9 (2021) 32.
\bibitem{Neil} R. Bluhm, V.A. Kosteleck\'y, N. Russell, Phys. Rev. Lett. 82 (1999) 2254-2257.
\bibitem{Manoel} R. Casana, M.M. Ferreira Jr., R.V. Maluf, F.E.P. dos Santos, Phys. Lett. B 726 (2013) 815.
\bibitem{AntAnomaly} A.P.B. Scarpelli, T. Mariz, J.R. Nascimento, A.Yu. Petrov, Int. J. Mod. Phys. A 31 (2016) 1650063.
\bibitem{Zonghao} V.A. Kosteleck\'y,  Z. Li, Phys. Rev. D 99 (2019) 056016.
\bibitem{Gazzola} A.R. Vieira, J.C.C. Felipe, G. Gazzola, M. Sampaio, Eur. Phys. J. C 75 (2015) 338. 
\bibitem{Shapiro} M. Asorey, E.V. Gorbar, I.L. Shapiro, Class. Quant. Grav. 21 (2003) 163-178.
\bibitem{BertlmannPLB} R. A. Bertlmann, E. Kohlprath, Phys. Lett. B 480 (2000) 200-206.
\bibitem{Leo} L.A.M. Souza, M. Sampaio, M.C. Nemes, Phys. Lett. B 632 (2006) 717-724. 
\bibitem{Orimar2} O.A. Battistel, G. Dallabona, Eur. Phys. J. C 45 (2006) 721.
\end{thebibliography}
\end{document}